\documentclass[aps,pra, twocolumn,groupedaddress]{revtex4-2}

\usepackage{graphicx}% Include figure files
\usepackage{dcolumn}% Align table columns on decimal point
\usepackage{bm}% bold math

\usepackage{amsmath}
\usepackage{gensymb}
\usepackage{amssymb}

\begin{document}

%\preprint{}

\title{Fast-speed algorithm to compute tight focusing of laser beams:
The effectiveness of circularly polarized vortex beam series as a mathematical basis}


\author{Stepan Boichenko}
\email[]{ste89@yandex.ru}
%\thanks{}
\affiliation{Irkutsk Branch of the Institute of Laser Physics of the Siberian Branch
of the Russian Academy of Sciences,
 Lermontov str. 130a, 664033 Irkutsk, Russian Federation}


\date{\today}

\begin{abstract}
We suggest a time-effective algorithm to calculate tight focusing of a
collimated continuous-wave laser beam  with an arbitrary cross-section
light vector distribution by a high-aperture microscope objective
into a planar microcavity.
This algorithm is based on the mathematical fact that any beam can be
decomposed into a superposition---either finite or infinite---of circularly polarized
vortex vector beams, which allows one to factorize focal field into two parts,
one of which depends only on distance coordinates $\rho$ and $z$ and
the other one only on an azimuth $\varphi$ in cylindrical coordinates.
We compare the suggested algorithm with that based on the direct use
of the double-integral Richards-Wolf method and demonstrate
that the former is at least 5 times faster for single-point computations
and at least two orders faster for typical focal-region computations.
\end{abstract}

\maketitle

\section{Introduction}
\label{intro}

A variety of microscopy techniques exploiting tightly focused laser beams is
widely used nowadays. These techniques are laser-scanning fluorescent microscopy
\citep{sick2000orientational, chizhik2011optical, foldes2003laser, pawley2006handbook},
STED microscopy \citep{hell1994breaking, vicidomini2018sted, blom2017stimulated},
multiphoton excitation microscopy \citep{larson2011multiphoton, konig2000multiphoton},
and some others \citep{novotny2012principles}.
In many researches, it is necessary to evaluate the focal-region light field
formed by tight focusing of a laser beam by a microscope objective and in
some cases it is desirable to know precisely the focal-region light field distribution
\citep{debarre2004quantitative, novotny2001longitudinal, chizhik2011optical, dolan2014complete}.
For this, one needs to have a tool to calculate theoretically such a distribution,
knowing entrance laser beam
light field distribution and microscope objective parameters.

A beam can be focused either in free space---that means an inhomogeneity-free
medium with continuous refractive index, not vacuum---or under
refractive-index-discontinuous conditions, for example, into an optical microcavity
\citep{rigneault1998resonant,khoptyar2008tight,wang2022monitoring},
below or above a glass/air interface \citep{sick2000orientational},
near a microcrack \citep{kovalets2022toward}, inside an optical nanoparticle
\citep{dolan2014complete}, and near other stochastic structures.
The task of laser beam focusing simulation in free space, near a planar
interface or in a planar microcavity is
completely solvable analytically and field values can be calculated numerically with nearly
unlimited precision, the only limitations of which
are concerned to the uncertainties in the microscope objective parameters,
microcavity mirrors, and entrance beam characteristics. Although such
uncertainties are inalienable, they can be mitigated in any
experiment to minimize differences between simulations and experimental
measurements.
In contrary, precise simulation of laser field focusing near optical
nanoparticles, microcracks, and other
stochastic structures is nearly impossible due to the lack of knowledge
of their geometrical forms.
But even when the forms can be defined,  the complexity of the task
of calculating a focal field near such a structure
is normally orders of magnitude greater than that of near-planar-structure
focusing and it is rarely solved in practice.
For this, the task of in-planar-microcavity laser beam tight focusing can
be considered as the most possible generalized
precisely solvable task and we will explore it here. The free-space tight
focusing and near-planar-interface tight focusing are particular
cases of this task.

Richards-Wolf theory \citep{richards1959electromagnetic}  is a key
universal tool to calculate tight focusing of laser
beams and, in principle, it allows one to calculate tight focusing of any beam.
Although the original method developed by Richards and Wolf was
designed for free-space focal field calculations,
it can be adapted to calculate near-planar-interface focusing without
severe changes in the approach \citep{khoptyar2008tight}.
The critical point is that  the original method includes double integration,
whereas  in most cases it is possible
to reduce focal field calculation procedure down to single integration
\citep{novotny2012principles}, which allows one to save a lot of calculation time.
Such a reduction can normally be done for any entrance beam and
attempts to develop the respective generalized
algorithm for  some families of entrance beams have been made.
So, tight focusing of non-vortex radially and azimuthally polarized laser
beams into planar microcavities was considered
by Meixner's group \citep{khoptyar2008tight} and the suggested algorithm
can be readily applied to this family of beams
but it cannot be applied, for example, to a vortex beam.
Free-space focusing of vortex laser beams was considered in 
\citep{zhan2006properties} but
the suggested algorithm cannot be applied to in-microcavity focusing
because of the mutual affection of a radial and azimuthal components
of a vortex beam. In fact, it had not been developed
any generalized single-integral focal field calculation algorithm for an
arbitrary entrance beam till present,
to the best of our knowledge, and in most cases one has to develop it
for a given entrance beam individually.
Meanwhile, for some types of entrance beams it  might be a complicated
time-consuming problem, on the one hand. On the other hand, a
generalized algorithm would save some time
anyway even for entrance beams, for which Richards-Wolf method
can be adapted easily.
In the present investigation we search for such an algorithm,  explore mathematical
basis for this, and, finally, suggest an algorithm.

The paper is organized as follows. In Sec.~\ref{CommonTheory} we describe the
problem and express the Richards-Wolf method in brief.
In Sec.~\ref{GenAlg8547} we search for a generalized mathematical structure of
an entrance beam and find a suitable structure, than adapt transfer-matrix method
to calculate tightly focused light field inside a planar microcavity.
In the next step, a generalized single-integral algorithm is described schematically
and the problem of an entrance beam expressed as an infinite Fourier series
is explored.
In Sec.~\ref{Computations4527} we test the suggested algorithm for its speed
on some basic entrance beams and a slowly convergent infinite-series beam
as representative examples.

\section{Focal field calculation: Common theory}
\label{CommonTheory}
\begin{figure}
\includegraphics{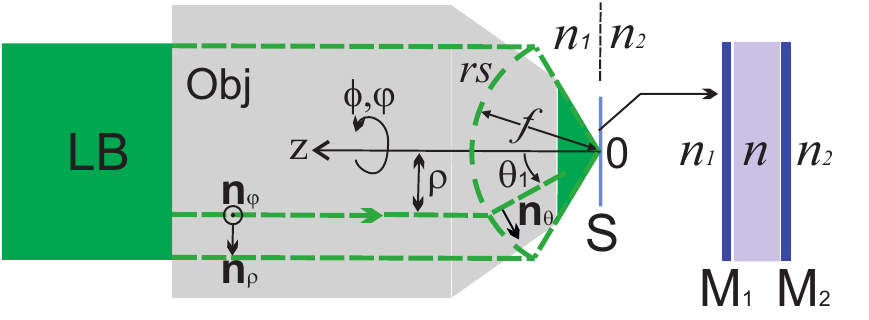}
\centering
\caption{A focusing scheme. A laser beam (LB) enters a
microscope objective (Obj) to be focused onto a sample (S),
which is assumed to be a planar microcavity. The objective
medium refractive index is assumed to be equal to $n_{1}$,
the microcavity core layer index $n$, and the right-hand side
medium index $n_2$. $ M_1$ and $M_2$ are the
microcavity mirrors. The vectors $\mathbf{n}_i$ with 
$i = \rho, \varphi, \theta$ are unit vectors of the respective
curvilinear coordinate systems; $\rho$, $\theta$, $\phi$,
$\varphi$, and $z$ are coordinates; \textit{rs} is the reference
sphere, and $f$ is the objective focal distance.}
\label{FocusScheme}
\end{figure}

Let us consider a laser beam propagating toward a microscope objective
pupil as shown in Fig.~\ref{FocusScheme} (here and below we will consider
Richards-Wolf method as adapted by Novotny and Hecht and described
in \citep{novotny2012principles}).
We assume that a beam is focused by a microscope objective into a
planar microcavity and we want to find light field in a given layer of interest
inside the microcavity (core layer).
The objective medium (left-hand side medium in the figure) refractive
index is assumed to be equal to $n_{1}$,
the microcavity core layer index $n$, and the right-hand side
medium index $n_2$.
It is assumed that the microscope objective satisfies the Abbe sine
condition, which means that
a focused laser beam with a plane-wave-like wavefront is converted
into a converging spherical wave and the entrance beam wavefront is
projected onto a sphere of radius $f$ (the reference sphere).

To calculate the focal-region light field, knowing the entrance beam field, we have
to define a coordinate system, in which the fields will be expressed.
The present focusing model assumes that (i) partial rays of the laser beam are
inclined at the reference sphere to be
directed to the geometrical focus and (ii) light field in a focal-region point is formed
by a superposition of fields
propagating from all points of the reference sphere, which means that one needs
to integrate the reference-sphere
field to obtain the focal-region one. As a laser beam and microscope objective
form a rotationally symmetric system,
it is straightforward to exploit cylindrical coordinates. However, the reference
sphere---being a part of a spherically
symmetric system---imposes spherical coordinates. To this end, we have to exploit
a cylindrical coordinate system with inclusion of spherical coordinates.

As shown in Fig.~\ref{FocusScheme}, we take the coordinate origin in the geometrical
focus of the objective and the objective axis $z$ is directed toward the entrance laser
beam. Note that a focal-region light field can be concentrated not exactly in the
geometrical focus (at $z=0$) but leftward (positive z's) or rightward
(negative z's) due to the microcavity refraction, but this fact does not prohibit keeping
the coordinate origin in the geometrical focus. Further, we introduce cylindrical coordinates
$(\rho,\varphi, z)$, where $\rho$ is the distance from
a point to the longitudinal axis $z$ and $\varphi$ is the azimuth in the cylinder cross
section. To define the azimuth, we need to introduce Cartesian coordinates $x$ and
$y$ in the cross section. But we have no preferential
directions there and can take as $x$ and $y$ any pair of mutually orthogonal axes
that form a right-hand coordinate system in conjunction with the shown $z$-axis.
The choice of such a pair does not play any important role,
the only noticeable condition is that positive values of the azimuth $\varphi$
correspond to the counterclockwise direction as shown in the figure.

To calculate the focal-region field, we, first, need to define two spacial points: one in the focal region where we
calculate the field and another one on the reference sphere or in the entrance beam cross section, which is
virtually the same.
Below, focal-region points will be described by coordinates $\rho$, $\varphi$, and $z$, reference-sphere points
by a polar angle $\theta_1$ and azimuthal angle $\phi$, beam cross-section points by $\rho$, $\phi$, and $z$.
We exploit the same names for the longitudinal and distance coordinates in the beam cross-section and in the
focal region but these groups of coordinates will not appear in the same focusing equations and such a choice of names
will not lead to any ambiguity. We will exploit the following three polar angles: $\theta_1$ in the objective space
(refractive index $n_1$), $\theta$ inside the microcavity (refractive index $n$), and $\theta_2$ in the right-hand-side
semi-space (refractive index $n_2$). In this notation, the reference-sphere polar angle is $\theta_1$.
The entrance beam cross-section distance coordinate and reference-sphere polar
angle are linked by the relation $\rho = f \sin{\theta_1}$. Vectors $\mathbf{n}_{\rho}$ and $\mathbf{n}_{\varphi}$ are unit vectors
of the distance and azimuth coordinates in a cylindrical coordinate system, $\mathbf{n}_{\theta}$ is the unit vector
of the polar coordinate in spherical coordinates. $\mathbf{n}_{\varphi}$---being orthogonal to the plane in which a partial
ray is inclined during the conversion of the entrance beam into a spherical wave---is not affected by this transformation
and remains the same on the reference sphere as it is in the cross-section and $\mathbf{n}_{\rho}$ is converted into
$\mathbf{n}_{\theta}$.

We assume the entrance beam light field to be a plane-wave-like monochromatic wave propagating toward the longitudinal axis
\begin{equation}
\label{EntranceFieldGen}
\mathbf{E}_{eb} (\rho,\phi,z,t) = \mathbf{E}_0 (\rho,\phi) \mathrm{e}^{i k_1 z} \mathrm{e}^{i \omega t}
\end{equation}
with $t$ being time, $\omega$ the angular frequency, $k_1 = n_1 \omega / c$ the wavenumber, $c$ the vacuum speed of light.
The conversion of the beam into a converging spherical wave imposes the beam cross-section part of the field
$\mathbf{E}_0 (\rho,\phi)$ to be transformed into
	\begin{align}
		\label{FieldInclined}
		\mathbf{E}_{inc} (\theta_1, \phi) = \sqrt{\cos(\theta_1)} \{ [ \mathbf{E}_0  (\theta_1, \phi) \cdot
		\mathbf{n}_{\rho} ] \mathbf{n}_{\theta} 
		\nonumber \\
		+ [ \mathbf{E}_0  (\theta_1, \phi) \cdot \mathbf{n}_{\varphi} ] \mathbf{n}_{\varphi} \}.
	\end{align}
To obtain a light field value in a free-space focal-region point, we have to calculate a superposition
of partial secondary plane waves propagating from the reference sphere to the focal region:
	\begin{align}
		\label{AngularSpectrum}
		\mathbf{E}_{foc} (\mathbf{r}) = 
		- \frac{i k_1 f \mathrm{e}^{  i  k_1 f}}{2 \pi} 
		\int \displaylimits_{0}^{\theta_{\mathrm{max}}} \int \displaylimits_{0}^{2 \pi}
		\mathbf{E}_{inc} (\theta_1, \phi)
		\mathrm{e}^{  i  \mathbf{k}(\theta_1,\phi) \cdot \mathbf{r}}
		\nonumber \\
		\times \sin{\theta_1} \, \mathrm{d} \phi \, \mathrm{d} \theta_1
	\end{align}
with $\mathbf{r}=(x,y,z)$ being Cartesian radius-vector of a focal-region point, $\theta_{\mathrm{max}}$ being
the objective angular aperture, and
\begin{equation}
\label{wavevector}
\mathbf{k}(\theta_1,\phi) = k_1
\begin{pmatrix}
 \sin{\theta_1} \cos{\phi}\\
 \sin{\theta_1} \sin{\phi} \\
 \cos{\theta_1}
\end{pmatrix}
\end{equation}
a wavevector. This expression can be considered as an angular plane-wave spectrum representation of the focal field
composed by waves with the vector amplitudes $\mathbf{E}_{inc}(\theta_1,\phi) \sin{\theta_1}$
and wavevectors $\mathbf{k}$.
To obtain in-microcavity focal field, we need to take into account action of such a cavity on
the field~(\ref{AngularSpectrum}).
As this field is represented by a superposition of mutually independent plane waves, we can readily explore
refraction and reflection
of each partial wave separately and then integrate cavity-converted waves to obtain in-cavity focal field.
Doing this, we can write in-cavity focused field as
	\begin{align}
		\label{FieldIntegral}
		\mathbf{E}_{foc} (\mathbf{r}) = - \frac{i k_1 f \mathrm{e}^{  i  k_1 f}}{2 \pi} 
		\int \displaylimits_{0}^{\theta_{\mathrm{max}}} \int \displaylimits_{0}^{2 \pi}
		\widehat{MC} \big [ \mathbf{E}_{inc} (\theta_1, \phi) 
		\nonumber \\
		\times \mathrm{e}^{  i  \mathbf{k}(\theta_1,\phi)
		\cdot \mathbf{r}}  \sin{\theta_1} \big ]
		\, \mathrm{d} \phi \,
		\mathrm{d} 				\theta_1 .
	\end{align}
The operator $\widehat{MC}$ represents the affection of the microcavity at
a partial plane wave. Details of its action will be considered in Subsection~\ref{EntranceBeam}.
Equation~(\ref{FieldIntegral}) in conjunction with Eqs.~(\ref{EntranceFieldGen}) and (\ref{FieldInclined}) is the
starting point in any collimated laser beam tight focusing simulation.

\section{Generalized single-integral algorithm}
\label{GenAlg8547}

In this section, we will explore the problem of a generalized single-integral algorithm to  calculate the laser beam tight focusing.
The major problem is that Eq.~(\ref{FieldIntegral}) contains an unknown vector amplitude $\mathbf{E}_0(\theta_1,\phi)$
and it might seem that one cannot calculate a focal field until this amplitude is completely defined.
However, if the amplitude can be partly defined, for example, decomposed in a Taylor series, Fourier series,
Laurent series or another generalized superposition, the focusing equations could be significantly simplified.
Hence, to develop a generalized algorithm of tight focusing field calculation, we need, first of all, to determine
a generalized structure of the entrance field.
Second, it is necessary to choose an optimal polarization basis to represent the entrance and focal fields.
To summarize, we should solve the problem of vector light field representation, which consists of the two
sub-problems: generalized representation of the scalar functions that describe light field vector components and
finding an optimal vector basis to decompose the field into three spatial components.

\subsection{Generalized mathematical structure of vector light fields}

Let us consider the vectorial representation of an entrance field.
The curvilinear unit vectors in Eq.~(\ref{FieldInclined}) are expressed in terms of the Cartesian 
unit vectors $\mathbf{n}_x$ and $\mathbf{n}_y$ as
%\begin{eqnarray}
\begin{subequations}
	\label{UnitVectors}
	\begin{gather}
	\mathbf{n}_{\rho} = \cos{\varphi} \, \mathbf{n}_{x} + \sin{\varphi} \, \mathbf{n}_{y}, \\
	\mathbf{n}_{\varphi} = - \sin{\varphi} \, \mathbf{n}_{x} + \cos{\varphi} \, \mathbf{n}_{y}, \\
	\mathbf{n}_{\theta} = \cos{\theta} \, \mathbf{n}_{\rho} - \sin{\theta} \, \mathbf{n}_{z}.
	\end{gather}
\end{subequations}
%\end{eqnarray}
The use of these vectors is imposed by the cylindrical symmetry of the microscope objective system
(this symmetry makes the use of Cartesian vectors completely improper),
but the problem of the cylindrical vectors is that they depend on the azimuthal angle and the dot products
$\mathbf{E}_0  (\theta, \phi) \cdot \mathbf{n}_{\rho}$ and $\mathbf{E}_0  (\theta, \phi) \cdot \mathbf{n}_{\varphi}$
cannon be directly integrated over $\phi$ in Eq.~(\ref{FieldIntegral}).
To perform the integration, we need to represent the integrated field in Cartesian or any other Cartesian-like
vector basis. Note that $\phi$-integration of radially and azimuthally polarized entrance beams with azimuth-independent
amplitudes, finally,
brings us back to radial and azimuthal unit vectors in focal region
\citep{novotny2012principles,khoptyar2008tight,boichenko2018theoretical}:
a radially polarized beam produces a radial and longitudinal focal components and an azimuthally polarized beam produces
an azimuthal component.
But in general, field amplitude is azimuth-dependent and the task cannot be solved in this way:
it will include the $\phi$-integration of functions $E_0^i (\theta_1,\phi) \cos{\phi}$ and
$E_0^i (\theta_1,\phi) \sin{\phi}$ ($i=\rho,\varphi$).
Such the integration requires an adequate series representation of entrance beam light field components.

\begin{figure}
	\includegraphics{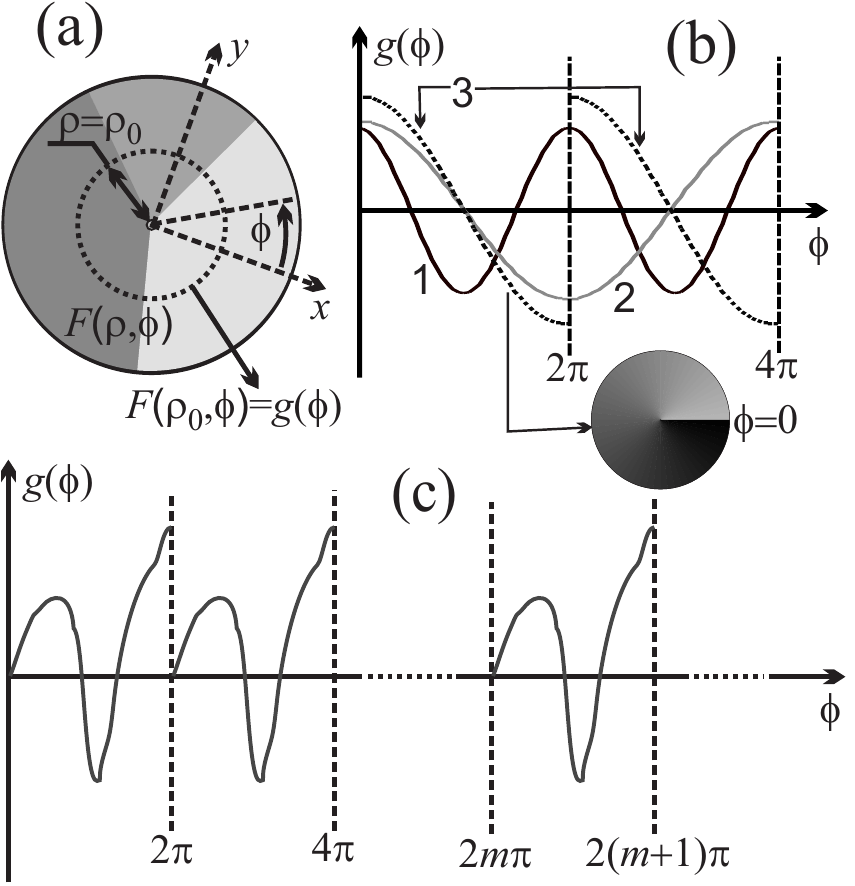}
	\centering
	\caption{An explanation of the light field azimuthal periodicity. (a) A laser beam cross-section scheme. (b) Examples
	of periodic functions. (c) An example of a generalized periodic function.}
	\label{PeriodicFunction}
\end{figure}

Among all possible series, Fourier series is the most attractive to expand the field components because, first,
they are closely connected to the functions $\cos{\phi}$ and $\sin{\phi}$ and, second, the field components
are inherently $\phi$-periodic functions with the period equal to $2 \pi$.
Some examples from practice may put into question this periodicity and Fig.~\ref{PeriodicFunction}
explains it in detail.
Figure~\ref{PeriodicFunction}a shows a laser beam cross-section. The axes $x$ and $y$ are Cartesian coordinate
axes, which can be chosen arbitrarily, $\phi$ is the azimuth angle and $\rho$ the distance coordinate in a cylindrical
coordinate system. We will consider a function $F(\rho,\phi)$ as a common field component.
Different grayscale tones schematically show that such a function is coordinate-dependent inside a laser
beam cross-section.
The angle $\phi$ is varied from 0 (it corresponds to $x$-direction) to $2 \pi$ and, obviously, a
$2 \pi$ term to an azimuth corresponds to a full-circle around the cylinder axis,
which guarantees the periodicity condition $F(\rho,\phi + 2 \pi) = F(\rho,\phi)$.
However, there exist some light beams---for example, fractional vortex beams
\citep{basistiy2004synthesis,wen2020observation}---described by aperiodic functions.
An example of such a beam can be expressed mathematically as $\mathbf{E}_0(\rho,\phi) = A(\rho)
\mathrm{e}^{i \phi /2} \mathbf{n}_x$ with $A(\rho)$ being a scalar amplitude and the azimuth depandence
is described by a function with the period of $4 \pi$ instead of $2 \pi$.
Figure~\ref{PeriodicFunction}b displays some examples of periodic functions to settle this point.
We denote a function $F(\rho,\phi)$ on a circle with the radius $\rho_0$ as  $g(\phi)$:
$F(\rho=\rho_0,\phi)=g(\phi)$. Next, a function $g(\phi)$ is displayed in rectangular coordinates
as shown in Figure~\ref{PeriodicFunction}b. The azimuth range is spread from $[0...2 \pi]$
to $[0...4 \pi]$ to explore the property of periodicity.
Three different functions are displayed. Curve $1$ corresponds to the function $g(\phi)=\cos{\phi}$ and
its period is equal to $2 \pi$. Curve $2$ corresponds to the function $g(\phi)=\cos(\phi/2)$
and its period is equal to $4 \pi$, which leads to the problem that the function takes on
different values in the ranges $[0...2 \pi]$ and $[2 \pi...4 \pi]$, while $(\rho_0, \phi + 2 \pi)$
corresponds to the same spatial point as $(\rho_0, \phi)$ and light field components
are single-valued functions of spatial coordinates.
The solution to this problem is demonstrated by curve $3$ and the respective inset.
Curves $2$ and $3$ correspond to the same function (with different amplitudes for the sake of
discernibility) in the range $[0...2 \pi]$, but at $\phi=2\pi$ curve $3$ has a jump instead of
going continuously according to the function $\cos(\phi/2)$.
So, if a light field component is described by a $\phi$-aperiodic function, the aperiodicity
indicates not the component's ambiguity but its discontinuity at $\phi=0$ (or $2 \pi$).
That means that we can (mathematically) spread the azimuth range from $[0...2\pi]$
to $[0...\infty]$ and a $g$-function will be periodic (with or without jump) as shown in
Fig.~\ref{PeriodicFunction}c. This conclusion is not valuable for direct expression of a light fields
in terms of mathematical functions, but crucially important for the problem of Fourier series
expansion possibility: it guarantees that a light field can be expressed mathematically
as a Fourier series in the range $\phi=[0...2\pi]$.

Now, we will define the vector basis to express light fields.
A Fourier series can be expressed either in terms of real functions $\sin(n\phi)$ and $\cos(n\phi)$
or in terms of complex functions $\exp( i n \phi)$, where $n$ is an integer.
If we use the former approach, the need to integrate terms $\sin(n\phi) \sin(\phi)$, $\sin(n\phi) \cos(\phi)$,
$\cos(n\phi) \sin(\phi)$, and $\cos(n\phi) \cos(\phi)$ in Eq.~(\ref{FieldIntegral}) will appear.
The latter approach is easier itself than the former one due to the uniformity of the basic functions 
and if the vectors $\mathbf{n}_{\rho}$ and $\mathbf{n}_{\varphi}$ are expressed in terms of
$\exp(\pm i \varphi)$, the complex-function Fourier series becomes even more suitable
because terms like $\exp(\pm i \varphi) \exp( i n \phi)$ possess the form $\exp(i m \phi)$ with
$m = n \pm 1$ and can be integrated directly.
To obtain a vector Fourier series, we should express the cylindrical vectors through the complex exponents.
Substituting expressions $\cos{\phi} = 0.5[\exp(i \phi) + \exp(-i \phi)]$ and
$\sin{\phi} = -0.5 i [\exp(i \phi) - \exp(-i \phi)]$ into Eqs.~(\ref{UnitVectors}), we obtain
\begin{subequations}
\label{CVBtoCircPol}
\begin{gather}
\mathbf{n}_{\rho} = 2^{-0.5} (e^{- i \varphi} \mathbf{n}_{+} + e^{ i \varphi} \mathbf{n}_{-}), \\
\mathbf{n}_{\varphi} = -2^{-0.5} i  (e^{- i \varphi} \mathbf{n}_{+} - e^{ i \varphi} \mathbf{n}_{-})
\end{gather}
\end{subequations}
with 
\begin{equation}
\label{Circ_in_Cart}
\mathbf{n}_{\pm} =2^{-0.5} (\mathbf{n}_{x} \pm i \mathbf{n}_{y})
\end{equation}
being unit vectors of a circular polarization basis.
The reverse transform reads as
\begin{subequations}
\label{CircPol}
\begin{gather}
\mathbf{n}_+ = 2^{-0.5} e^{i \varphi} (\mathbf{n}_{\rho} + i \mathbf{n}_{\varphi}), \\
\mathbf{n}_- = 2^{-0.5} e^{-i \varphi} (\mathbf{n}_{\rho} - i \mathbf{n}_{\varphi}).
\end{gather}
\end{subequations}
In the next step, we will explore light field Fourier series representation in a circular polarization
vector basis.

Any light field with a total angular momentum defined by a topological charge $m$,
propagating toward $z$-direction can be expressed as a complex superposition of the following two fields:
\begin{subequations}
\label{EntranceField}
\begin{align}
	\mathbf{E}_+(\rho,\phi,z) = \exp[i (m-1) \phi] A_+(\rho,z) \mathbf{n}_+
	\nonumber \\
	+ \exp(i m \phi) A_z^+(\rho,z) \mathbf{n}_z, 
\end{align}
\begin{align}
	\mathbf{E}_-(\rho,\phi,z) = \exp[i (m+1) \phi] A_-(\rho,z) \mathbf{n}_-
	\nonumber \\
	+ \exp(i m \phi) A_z^-(\rho,z) \mathbf{n}_z,
\end{align}
\end{subequations}
with the amplitudes
\begin{subequations}
\label{EntranceFieldAmplitudes}
\begin{gather}
	A_{\pm}(\rho,z) = \int \displaylimits _{0}^{\infty} Q_{\pm}(q) J_{m \mp 1}(q \rho)\exp[i k(q) z] \mathrm{d}q,
	\label{EntranceFieldAmplitudes1}\\
	A_z^{\pm}(\rho,z) = \mp i \int \displaylimits _{0}^{\infty} [q/k(q)] Q_{\pm}(q)
	J_{m} (q \rho)\exp[i k(q) z] \mathrm{d}q, \label{EntranceFieldAmplitudes2}
\end{gather}
\end{subequations}
where $k(q) = \sqrt{(2 \pi n / \lambda)^2 - q^2}$ is a longitudinal wavenumber, $q$ a transverse extinguish
parameter, $Q_{\pm}(q)$ functions of the parameter $q$, and $m$ an integer \citep{boichenko2021orientational}.
The parameter $q$ is varied up to infinity in the Bessel beam field representation in
Eqs.~(\ref{EntranceFieldAmplitudes}), but a laser beam to be paraxial (as we are considering such a beam), functions
$Q(q)$ must be non-vanishing only
in a narrow range of $q$-values and these values shoul be of the order of $1/f$.
Although Bessel beams with high $q$-values are non-diffractive like those with low $q$-values,
their plane-wave angular spectrum representation is formed by waves with large transverse components of
the wavevector \citep{mcgloin2005bessel} and they cannot be tightly focused and considered as paraxial.
For this, we have the approximation
$q/k(q) \propto \lambda / f \propto 10^{-3}$ as light wavelength $\lambda$ is of the order of hundreds of
nanometers and the focal length $f$ is of the orders of millimeters and the longitudinal component in
Eq.~(\ref{EntranceField}) can be neglected. Further, the integral in Eq.~(\ref{EntranceFieldAmplitudes1})
can be approximated as
\begin{equation}
 \int \displaylimits _{0}^{\infty} Q_{\pm}(q) J_{m \pm 1}(q \rho)\exp[i k(q) z] \mathrm{d}q
\approx G_m^{\pm}(\rho) \exp(i k z),
\end{equation}
where the factor G, in general, weakly depends on $z$ but, first, this dependence is weak itself and, second,
$z$-coordinate corresponding to the reference sphere position must be defined and assigned in calculations,
which allows one to assume the function G to be dependent on $\rho$-coordinate only.
Thus, a light field of a paraxial laser beam with a topological charge $m$ can be expressed as
\begin{equation}
	\label{FieldmMoment}
	\mathbf{E}_m (\rho,\phi,z) = \mathrm{e}^{i m \phi} [\mathrm{e}^{- i \phi} G_m^+(\rho) \mathbf{n}_+
	+\mathrm{e}^{ i \phi} G_m^-(\rho) \mathbf{n}_-] \mathrm{e}^{ i k z}.
\end{equation}
To obtain a generalized decomposition of a light field of an arbitrary paraxial beam, we should summarize
fields (\ref{FieldmMoment}) over all the topological charges: $m = [-\infty...\infty]$.
Summarizing them and substituting $G_{m+1}^+(\rho)$ by $G_{m}^+(\rho)$ and $G_{m-1}^-(\rho)$
by $G_{m}^-(\rho)$, we obtain the generalized expression for an entrance field $\mathbf{E}(\rho,\phi)$
in Eq.~(\ref{EntranceField})
\begin{equation}
\label{EntranceFieldAppr}
\mathbf{E}_0(\rho,\phi) =  E_+(\rho,\phi) \mathbf{n}_+ + E_-(\rho,\phi) \mathbf{n}_- 
\end{equation}
with % Precise Maxwell's eqs. solutions!!!
\begin{subequations}
\label{EntranceComponentsAppr}
\begin{gather}
{E}_+(\rho,\phi) = \sum_{m=-\infty}^{\infty} G_m^+(\rho) \exp (i m \phi), \\
{E}_-(\rho,\phi) = \sum_{m=-\infty}^{\infty}  G_m^-(\rho) \exp (i m \phi).
\end{gather}
\end{subequations}
This equation is a complex-function Fourier series representation of an entrance field in a circular polarization
vector basis. Such a representation is pretty generalized and can be exploited further to develop a generalized
algorithm of tight focusing calculation.

\subsection{Entrance laser beam light field: The focusing process}
\label{EntranceBeam}
Now, we can directly integrate the decomposition of an entrance field~(\ref{EntranceFieldAppr})
to solve the focusing equation.
As all the terms in Eq.~(\ref{EntranceFieldAppr}) are mutually independent, they should be
integrated separately and a generalized partial field to be integrated is expressed as
\begin{equation}
\label{FieldPartialCirc}
\mathbf{E}^{\pm}_m (\rho,\phi) = G_m^{\pm}(\rho) \exp (i m \phi) \mathbf{n}_{\pm}.
\end{equation}
Then to obtain a total focal field, we should summarize focused partial fields over all $m$'s appearing in
Eq.~(\ref{EntranceFieldAppr}).
To treat a partial field~(\ref{FieldPartialCirc}) at the reference sphere, we should represent it 
in terms of cylindrical vectors, using Eqs.~(\ref{CircPol}) and  it  takes the form
\begin{equation}
\label{FieldPartialCyl}
\mathbf{E}_m^{\pm} (\rho,\phi,z) = 2^{-0.5} G_m^{\pm}(\rho) \exp [i (m \pm 1) \phi] (\mathbf{n}_{\rho} \, \pm \,
i \, \mathbf{n}_{\varphi}).
\end{equation}
Next, we introduce a new function $F_m^{\pm}(\theta_1) = G_m^{\pm}(f \sin{\theta_1})$ for the sake of brevity and
by combining Eqs.~(\ref{UnitVectors}) and (\ref{FieldInclined})
the inclination of the field~(\ref{FieldPartialCyl}) at the reference sphere gives
\begin{align}
\label{FieldCPinclined}
\mathbf{E}_{inc}(\theta_1, \phi) = 2^{-0.5} \sqrt{\cos{\theta_1}} F_m^{\pm}(\theta_1) \exp[i (m \pm 1) \phi]
\nonumber \\
\times [\cos{\theta_1} \, \mathbf{n}_{\rho} \pm i \,  \mathbf{n}_{\varphi} - \sin{\theta_1} \,
\mathbf{n}_z].
\end{align}
The index $m$ at the reference-sphere field was omitted as this field is an intermediate entity.

Next, we will consider the action of the microcavity on the light field.
Substituting the field~(\ref{FieldCPinclined}) into the focusing equation~(\ref{AngularSpectrum}) and
denoting a scalar amplitude as $A(\theta_1,\phi) = 2^{-0.5} \sqrt{\cos{\theta_1}} \sin{\theta_1}
F_m^{\pm}(\theta_1) \exp[i (m \pm 1) \phi]$, we can consider light field of a partial plane wave as
\begin{align}
\label{PartialPW}
\mathbf{E}_{pw}(\mathbf{r};\theta_1,\phi) = A(\theta_1,\phi)
[(\cos{\theta_1} \, \mathbf{n}_{\rho} 
\nonumber \\
- \sin{\theta_1} \,
\mathbf{n}_z) \pm i \,  \mathbf{n}_{\varphi}] \mathrm{e}^{i \mathbf{k}(\theta_1,\phi) \cdot \mathbf{r}}
\end{align}
or in the form
\begin{equation}
\label{SPsuperposition}
\mathbf{E}_{pw}(\mathbf{r};\theta_1,\phi) = \mathbf{E}_{pw}^{(s)}(\mathbf{r};\theta_1,\phi) +
\mathbf{E}_{pw}^{(p)}(\mathbf{r};\theta_1,\phi)
\end{equation}
with
\begin{subequations}
	\label{sp-waves}
	\begin{gather}
	\mathbf{E}_{pw}^{(s)}(\mathbf{r};\theta_1,\phi) = \pm i A(\theta_1,\phi)
	\,  \mathbf{n}_{\varphi} \mathrm{e}^{i \mathbf{k}(\theta_1,\phi) \cdot \mathbf{r}}, \\
	\mathbf{E}_{pw}^{(p)}(\mathbf{r};\theta_1,\phi) = A(\theta_1,\phi)
	(\cos{\theta_1} \, \mathbf{n}_{\rho} - \sin{\theta_1} \, \mathbf{n}_z)
	\mathrm{e}^{i \mathbf{k}(\theta_1,\phi) \cdot \mathbf{r}},
	\end{gather}
\end{subequations}
where indices $s$ and $p$ denote $s$-polarized (or transverse-electric) and $p$-polarized
(or transverse-magnetic) waves, respectively. To rewrite Eq.~(\ref{PartialPW}) in the form~(\ref{SPsuperposition}),
we exploited the fact that an azimuthal polarization corresponds to
$s$-polarized waves and a radial polarization to $p$-waves  in the angular spectrum representation.
Exploiting Eq.~(\ref{PartialPW}), we can explore microcavity action on partial plane waves.
Denoting microcavity operator as $\widehat{MC}$ and applying it to Eqs.~(\ref{sp-waves}),
one obtains
\begin{subequations}
	\label{MCsp-waves}
	\begin{align}
	\widehat{MC} \big [ \mathbf{E}_{pw}^{(s)}(\mathbf{r};\theta_1,\phi) \big ] = \pm i \,
	T_{\varphi}(\theta_1, z) A(\theta_1,\phi)
	\,  \mathbf{n}_{\varphi} \mathrm{e}^{i \mathbf{k}_{\perp} (\theta_1,\phi) \cdot \mathbf{r}_{\perp}},
	\end{align}
	\begin{align}
	\widehat{MC} \big [ \mathbf{E}_{pw}^{(p)}(\mathbf{r};\theta_1,\phi) \big ] = A(\theta_1,\phi)
	[ T_{\rho}(\theta_1, z) \cos{\theta} \, \mathbf{n}_{\rho}
	\nonumber \\
	- T_{z} (\theta_1, z) \sin{\theta} \, \mathbf{n}_z ]
	\mathrm{e}^{i \mathbf{k}_{\perp}(\theta_1,\phi) \cdot \mathbf{r}_{\perp}}.
	\end{align}
\end{subequations}
The symbol $\perp$ denotes vector components orthogonal to $z$-direction.
In this equation the angle $\theta$ depends on the angle $\theta_1$
and this dependence is defined from Snell's law as $n \sin{\theta} = n_1 \sin{\theta_1}$.
Here and below we assume the presence of such the dependence, although normally do not display it in
equations apparently. $T_i(\theta_1,z)$ ($i = \rho, \varphi, z$) are in-cavity interference coefficients.
Due to the in-cavity multiple reflections a plane wave propagating toward $z$-direction transforms 
into a superposition of two waves, one of which propagates toward $z$-direction and the other one
in opposite direction and this fact is reflected in $z$-dependence of the interference coefficients.
To solve the focusing equation, we need to calculate these coefficients from microcavity parameters.

\begin{figure}
	\includegraphics{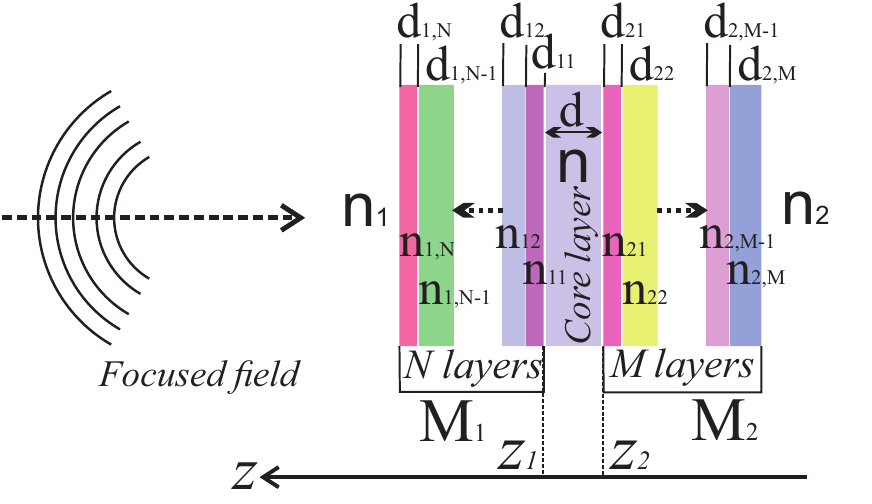}
	\centering
	\caption{The detailed structure of a microcavity shown in Fig.~\ref{FocusScheme}. $M_1$ and $M_2$ are
	planar layered mirrors, the objective medium semi-space refractive index is equal to $n_1$, and the right-hand
	side medium refractive index is equal to $n_2$. Focused laser field is calculated in a core layer and left
	and right bounds of the core layer are described by $z_1$ and $z_2$ coordinates, respectively.
	Mirror $M_1$ consists of $N$ layers and mirror $M_2$ of $M$ layers; $j^{th}$ layer of an $i^{th}$ mirror
	is characterized by a refractive index $n_{i,j}$ and thickness $d_{i,j}$.}
	\label{Cavity}
\end{figure}

Figure~\ref{Cavity} displays the detailed structure of a microcavity under consideration.
Mirrors that form such a cavity can be considered as planar stratified media and described by the matrices
\begin{subequations}
\label{MirrorMatr}
\begin{gather}
M_1 =
\begin{pmatrix}
d_{11} & d_{12} & \ldots & d_{1,N}\\
n_{11} & n_{12} & \ldots & n_{1,N}
\end{pmatrix}, \label{Mirror1M} \\
 \label{Mirror2M}
M_2 = 
\begin{pmatrix}
d_{21} & d_{22} & \ldots & d_{2,M}\\
n_{21} & n_{22} & \ldots & n_{2,M}
\end{pmatrix},
\end{gather}
\end{subequations}
where $d_{ij}$ and $n_{ij}$ are the thickness and refractive index of $j^{th}$ layer of the $i^{th}$ mirror
as shown in Fig.~\ref{Cavity} with $i = 1,2$ being the index of a mirror and $j$ a layer's index:
$j  = 1 \dots N$ at $i=1$, $j=1 \dots M$ at $i=2$.
It is assumed that mirror $M_1$ consists of $N$ layers and mirror $M_2$ of $M$ layers.
Focused field is passed through mirror $M_1$ from the objective medium with the refractive index $n_1$ into the
core layer of a thickness $d$ and with the refractive index $n$, then multiply reflected from the two mirrors inside
the core layer to form the final focused field.
A partial plane wave undergoes the following three processes: (1) transmission from medium $1$ into
the core layer through mirror $M_1$, (2) reflection from mirror $M_2$, (3) reflection from mirror $M_1$.
The transmission process occurs only one time for a given plane wave and the reflections are multiple.
We will denote $t^{\alpha}_1(\theta_1)$ the transmission coefficients and $r^{\alpha}_i(\theta_1)$
the reflection coefficients with $\alpha = s,p$ being the wave polarization index and $i=1,2$ the mirror index.
Summarizing multiply reflected waves, we obtain the in-cavity interference coefficients
\begin{widetext}
\begin{equation}
T_{m}(\theta_1,z) = t_{1}^{\alpha_m}(\theta_1)
\frac{\exp [ -i k_z(\theta_1) z ] + s_m r_{2}^{\alpha_m}(\theta_1) \exp [ i k_z(\theta_1) (z - 2 z_2) ]}
{1 - r_{2}^{\alpha_m}(\theta_1) r_{1}^{\alpha_m}(\theta_1) \exp [ 2 i k_z(\theta_1) d ]}
\end{equation}
\end{widetext}
with  $m=\rho,\varphi,z$; $k_z(\theta_1) = k_0 n \cos[\theta(\theta_1)]$, $s_{\rho} = -1$,
$s_{\varphi} = 1$, $s_{z} = 1$; $\alpha_{\rho}, \alpha_{z}=p$, $\alpha_{\varphi} = s$.
Reflection and transmission coefficients of the mirrors can be directly calculated
by means of the transfer-matrix method. We will adapt this method from the textbook \citep{born2013principles}.
Following this way, the first step is to form two vectors (one dimensional arrays) of characteristic matrices from
the two mirror matrices (\ref{Mirror1M}) and (\ref{Mirror2M}):
\begin{subequations}
\begin{gather}
L_1^{\alpha} (\theta_1) =
\begin{pmatrix}
L_{11}^{\alpha} (\theta_1) & L_{12}^{\alpha} (\theta_1) & \ldots & L_{1,N}^{\alpha} (\theta_1)
\end{pmatrix}, \label{Mirror1LM} \\
 \label{Mirror2LM}
L_2^{\alpha} (\theta_1) = 
\begin{pmatrix}
L_{21}^{\alpha} (\theta_1) & L_{22}^{\alpha} (\theta_1) & \ldots & L_{2,M}^{\alpha} (\theta_1)
\end{pmatrix},
\end{gather}
\end{subequations}
with
\begin{widetext}
\begin{equation}
L_{jm}^{\alpha} (\theta_1) =
\begin{pmatrix}
\cos[\beta_{jm} (\theta_1) ] & (- i / \{ n_{jm}^{g_{\alpha}} \cos[\theta_{jm}(\theta_1)] \}  ) \sin{\beta_{jm}} \\
- i \, n_{jm}^{g_{\alpha}} \cos[\theta_{jm}(\theta_1)]  \sin{\beta_{jm}} & \cos[\beta_{jm}(\theta_1)]
\end{pmatrix},
\end{equation}
\end{widetext}
where $\beta_{jm}(\theta_1) = (2 \pi / \lambda) n_{jm} d_{jm} \cos[\theta_{jm}(\theta_1)]$ with $\lambda$
being laser light
vacuum wavelength; $\alpha = s,p$ the wave polarization index; $g_{\alpha}$ refractive index power of the 
respective wave: $g_s = 1$, $g_p = -1$; angles $\theta_{jm}$ are functions of the
angle of incidence $\theta_1$ defined from Snell's law: $n_1 \sin{\theta_1} = n_{jm} \sin[\theta_{jm}(\theta_1)]$.
To define the transmission coefficients $t_1^{s}$ and $t_1^p$, we need, first, to calculate the characteristic
matrix of the first mirror for the direction from medium 1 to the core layer of microcavity. This matrix is
calculated as a matrix product
\begin{equation}
\label{MatProdRev}
U_1^{\alpha}(\theta_1) =  L_{1,N}^{\alpha} (\theta_1)  L_{1,N-1}^{\alpha} (\theta_1) \ldots
L_{12}^{\alpha} (\theta_1)  L_{11}^{\alpha} (\theta_1)
\end{equation}
and the transmission coefficients are expressed as
\begin{equation}
t_1^{\alpha} (\theta_1) = \frac{2 v_{\alpha} \cos{\theta_1}}{ f_1(\theta_1)+ f_2(\theta_1)}
\end{equation}
with $v_s = n_1$ and $v_p = 1 / n$,
\begin{equation}
f_1(\theta_1) = \{ [U_1^{\alpha}(\theta_1)]_{11} +
[U_1^{\alpha}(\theta_1)]_{12} n^{g_{\alpha}}
\cos[\theta(\theta_1)]  \} n_1 ^{g_{\alpha}} \cos{\theta_1} 
\end{equation}
and
\begin{equation}
f_2(\theta_1) = [U_1^{\alpha}(\theta_1)]_{21} +
[U_1^{\alpha}(\theta_1)]_{22} n ^{g_{\alpha}} \cos[\theta(\theta_1)]. 
\end{equation}
Similarly, to calculate  refractive index of the $j^{th}$ mirror in the direction from the core layer into
the $j^{th}$ semi-space, we need to find characteristic matrix of the $j^{th}$ mirror as a product
\begin{equation}
V_j^{\alpha}(\theta_1) =  L_{j,1}^{\alpha} (\theta_1)  L_{j,2}^{\alpha} (\theta_1) \ldots
L_{j,Q_j-1}^{\alpha} (\theta_1)  L_{j,Q_j}^{\alpha} (\theta_1)
\end{equation}
with $Q_1=N$ and $Q_2=M$. The refractive indices are expressed as
\begin{equation}
\label{MatProdDir}
r_j^{\alpha} (\theta_1) = \frac{g_1(\theta_1)  - g_2(\theta_1)}{ g_1(\theta_1)  +  g_2(\theta_1)}.
\end{equation}
with
\begin{equation}
g_1(\theta_1) = \big \{ [V_j^{\alpha}(\theta_1)]_{11} +
[V_j^{\alpha}(\theta_1)]_{12} n_j^{g_{\alpha}}
\cos{\theta_j} \big \} n ^{g_{\alpha}} \cos{\theta}
\end{equation}
and
\begin{equation}
g_2(\theta_1) =  [V_j^{\alpha}(\theta_1)]_{21} +
[V_j^{\alpha}(\theta_1)]_{22} n_j ^{g_{\alpha}} \cos{\theta_j}.
\end{equation}
Here, angle $\theta_j$ is equal to $\theta_1$ for $j=1$ and $\theta_2(\theta_1)$ for $j=2$ defined
from Snell's law as $n_2 \sin[\theta_2(\theta_1)] = n_1 \sin{\theta_1}$.
Note that in matrix products (\ref{MatProdRev}) and (\ref{MatProdDir}) matrices $L_{j,m}^{\alpha} (\theta_1)$
are multiplied in opposite orders and matrices $V_1^{\alpha}(\theta_1)$ and $U_1^{\alpha}(\theta_1)$ are
not identical because $L_{j,m}^{\alpha} (\theta_1)$ and $L_{j,l}^{\alpha} (\theta_1)$ are not mutually
commutative in general.

Summarizing the results obtained above, we can derive a equation of microcavity action on a partial
plane wave~(\ref{PartialPW})
%\begin{widetext}
\begin{align}
\label{FieldInclCav}
\widehat{MC}[\mathbf{E}_{pw}(\mathbf{r};\theta_1,\phi) ] =
2^{-0.5} \sqrt{\cos{\theta_1}} F_m^{\pm}(\theta_1) 
 \nonumber \\
\times \exp[i (m \pm 1) \phi]  
 \, \big [ T_{\rho}(\theta_1,z) \cos{\theta} \, \mathbf{n}_{\rho}
\nonumber \\
\pm i \, T_{\varphi}(\theta_1,z) \, \mathbf{n}_{\varphi} -
T_{z}(\theta_1,z) \sin{\theta} \, \mathbf{n}_z \big ] \mathrm{e}^{i \mathbf{k}_{\perp}(\theta_1,\phi)
\cdot \mathbf{r}_{\perp}}.
\end{align}
%\end{widetext}
To integrate this field over the angles $\theta_1$ and $\phi$, we should represent it in terms of
$\mathbf{n}_{\pm}$ polarization vectors as
%\begin{widetext}
\begin{align}
\label{MicroCavCirc}
\widehat{MC}[ \mathbf{E}_{pw}(\mathbf{r};\theta_1,\phi) ] =
\big [  E_{inc}^{+,mc}(\theta_1,\phi;z) \mathbf{n}_+ 
\nonumber \\
+ E_{inc}^{-,mc}(\theta_1,\phi;z) \mathbf{n}_-
+ E_{inc}^{z,mc}(\theta_1,\phi;z) \mathbf{n}_z \big ] \mathrm{e}^{i \mathbf{k}_{\perp}(\theta_1
\phi) \cdot \mathbf{r}_{\perp}}.
\end{align}
%\end{widetext}
Substituting Eqs.~(\ref{CVBtoCircPol}) expressing a cylindrical vector beam basis through a circular one
into Eq.~(\ref{FieldInclCav}), we obtain the following expressions for the amplitudes in  Eq.~(\ref{MicroCavCirc}):
\begin{subequations}
\label{FieldIncComponents}
\begin{align}
E_{inc}^{+,mc}(\theta_1,\phi;z) = 0.5 \sqrt{\cos{\theta_1}} F_m^{\pm}(\theta) [T_{\rho}(\theta,z)
\cos{\theta} 
\nonumber \\
\pm T_{\varphi}(\theta,z)] \exp[ i (m \pm 1 - 1) \phi], 
\end{align}
\begin{align}
E_{inc}^{-,mc}(\theta_1,\phi;z) = 0.5 \sqrt{\cos{\theta_1}} F_m^{\pm}(\theta)
[T_{\rho}(\theta,z) \cos{\theta}
\nonumber \\
\mp T_{\varphi}(\theta,z)] \exp[ i (m \pm 1 + 1) \phi], 
\end{align}
\begin{align}
E_{inc}^{z,mc}(\theta_1,\phi;z) = - (1/\sqrt{2}) \sqrt{\cos{\theta_1}} F_m^{\pm}(\theta)
\nonumber \\
T_z(\theta,z) \sin{\theta} \exp [i (m \pm 1) \phi].
\end{align}
\end{subequations}
The vector $\mathbf{k}_{\perp}(\theta_1,\phi)$ is expressed by Eq.~(\ref{wavevector}) and the vector $\mathbf{r}$
takes the form $\mathbf{r}=(\rho \cos{\varphi}, \rho \sin{\varphi}, z)$ in cylindrical coordinates and
the dot product $\mathbf{k}_{\perp}(\theta_1,\phi) \cdot \mathbf{r}_{\perp}$ takes the form
$\mathbf{k}_{\perp}(\theta_1,\phi) \cdot \mathbf{r}_{\perp} = k_1 \rho \sin{\theta_1} \cos({\varphi - \phi})$.
Although we work in the core layer with the refractive index $n$ and polar angle $\theta$,
the in-plane wavevecror component $\mathbf{k}_\perp$ is conserved according to Snell's law,
which allows us to express it directly in terms of $n_1$ and $\theta_1$.
To integrate this field over the azimuthal angle, we will exploit the relationship
\begin{align}
\label{BessFuncIntegr4237}
\int \displaylimits_0^{2 \pi} \exp(i l \phi) \exp[ i k_1 \rho \sin{\theta_1} \cos(\varphi - \phi) ] \mathrm{d}\phi
\nonumber \\
= 2 \pi J_l(k_1 \rho \sin{\theta_1}) \exp[i (\varphi + \pi/2) l ]
\end{align}
valid for an integer $l$ with $J_l(k \rho \sin{\theta})$ being the $l^{th}$-order Bessel function of the first kind
\citep{novotny2012principles}.
Finally, partial focal field takes the form
%\begin{widetext}
\begin{align}
\label{FinalFocField}
\mathbf{E}_{m,\pm}^f (\rho, \varphi, z) = 0.5 i k_1 f \mathrm{e}^{  i  k_1 f}
\exp[i (\varphi + \pi / 2) (m \pm 1)]
\nonumber \\
\times [A^{m}_{+,\pm}(\rho,z) \mathrm{e}^{ - i (\varphi + \pi /2)}
\mathbf{n}_+ + A^{m}_{-,\pm}(\rho,z) \mathrm{e}^{ i (\varphi + \pi /2)} \mathbf{n}_-
\nonumber \\
+ A^{m}_{z,\pm}(\rho,z) \mathbf{n}_z]
\end{align}
%\end{widetext}
with
\begin{subequations}
\label{FieldIncComponentsIntegral}
\begin{align}
A_{+,\pm}^{m}(\rho,z) = \int \displaylimits_{0}^{\theta_{max}} \sqrt{\cos{\theta_1}} \sin{\theta_1}
 F_m^{\pm}(\theta_1)
\nonumber \\
\times J_{m \pm 1 -1 }(k_1 \rho \sin{\theta_1})
[T_{\rho}(\theta_1,z) \cos{\theta} \pm T_{\varphi}(\theta_1,z)] \mathrm{d} \theta_1,
\end{align}
\begin{align}
A_{-,\pm}^{m}(\rho,z) = \int \displaylimits_{0}^{\theta_{max}}  \sqrt{\cos{\theta_1}}  \sin{\theta_1} F_m^{\pm}(\theta_1)
\nonumber \\
\times J_{m \pm 1 +1 }(k_1 \rho \sin{\theta_1})
[T_{\rho}(\theta_1,z) \cos{\theta} \mp T_{\varphi}(\theta_1,z)]  \mathrm{d} \theta_1,
\end{align}
\begin{align}
A_{z,\pm}^{m}(\rho,z) = - \sqrt{2} \int \displaylimits_{0}^{\theta_{max}} \sqrt{ \cos{\theta_1}}
\sin{\theta_1} \sin{\theta} F_m^{\pm}(\theta_1)
\nonumber \\
\times J_{m \pm 1 }(k_1 \rho \sin{\theta_1}) T_z(\theta_1,z)  \mathrm{d} \theta_1 .
\end{align}
\end{subequations}
Now, we are able to compute programmatically partial focal fields from Eqs.~\ref{FieldIncComponentsIntegral},
knowing technical parameters of a microcavity, microscope objective, and entrance laser beam light field.
To obtain focal field formed by tight focusing of a given beam, we should properly combine
its partial fields and the next subsection is devoted to the details of such a procedure.

\subsection{Generalized single-integral algorithm: Formalized scheme}

Initially, we have an entrance laser beam with light field described by Eq.~(\ref{EntranceFieldGen}), vector amplitude 
of which can be presented in any polarization basis. In the first step, we need to represent it in a circular polarization
basis as in Eq.~(\ref{EntranceFieldAppr}) if it is presented in another one. Then the amplitudes $E_+(\rho,\phi)$ 
and $E_-(\rho,\phi)$ should be decomposed into complex-function Fourier series~(\ref{EntranceComponentsAppr}).
For azimuth-continued simple fields this series reduces to a finite sum, but for azimuth-discontinued and other
complicated fields
the series is infinite. An example of a mathematical function corresponding to an azimuth-continued but infinite-series field
might be, for instance, $\exp(\sin{\phi})$. Although light fields obeying Maxwell's equations and described by
this function barely exist, it explains the concept of azimuth-continued infinite-series fields. In case of infinite
series, we should assign the precision of the focal field we want to obtain, assess the respective number of
terms we need to reach this precision and then reduce the series to the respective finite sum.

So, mathematically, we can associate with amplitudes $E_+(\rho,\phi)$ and $E_-(\rho,\phi)$ a set of integer-valued
numbers $M$
\begin{equation}
\label{SetInteger}
M = \{ m_1, m_2 \ldots m_p \},
\end{equation}
which consists of $p$ elements and
light field radial amplitudes can be expressed as vectors
\begin{subequations}
\label{VectorAmpl746}
\begin{gather}
\mathbf{G}_+(\rho) = 
\big (
G^+_{m_1}(\rho), G^+_{m_2}(\rho) \ldots G^+_{m_p}(\rho)
\big ), \\
\mathbf{G}_- (\rho)=
\big (
G_{m_1}^-(\rho), G_{m_2}^-(\rho) \ldots G_{m_p}^-(\rho)
\big ).
\end{gather}
\end{subequations}
Further, introducing a vector of exponents
\begin{equation}
\label{ExpVector74}
\mathbf{e}(\phi) = 
\big (
\exp(i m_1 \phi), \exp(i m_2 \phi) \ldots \exp(i m_p \phi)
\big ),
\end{equation}
one can express amplitudes~(\ref{EntranceComponentsAppr}) as scalar products
\begin{subequations}
\label{AmpsAsSer}
\begin{gather}
{E}_+(\rho,\phi) = \mathbf{G}_+(\rho) \cdot  \mathbf{e}(\phi) , \\
{E}_-(\rho,\phi) = \mathbf{G}_-(\rho) \cdot  \mathbf{e}(\phi).
\end{gather}
\end{subequations}
Note that in this notation some components of vectors $\mathbf{G}_{\pm}(\rho)$ may vanish because
the set $M$ was introduced to contain both positive and negative polarization amplitudes, which are mutually
independent and a given vortex beam can possess only one polarization component.
To avoid vanishing components,
we could describe amplitudes $E_+$ and $E_-$ by different sets $M_+$ and $M_-$ but as
any entrance component induces all the three spatial components in the focal area,
the approach of one set is preferable.
The procedure of calculation of a partial focal field~(\ref{FinalFocField}) induced by an entrance field~(\ref{FieldPartialCirc})
at known microcavity mirror parameters and other technical settings is finely programmable
and field focusing task  can be assumed to be solved if a computer receives at the entrance field radial amplitudes.
Thus, an entrance field to be defined for a computer program, a user needs to set vectors $\mathbf{G}_+(\rho)$
and $\mathbf{G}_-(\rho)$ or---which would be better in practice---vectors $\mathbf{F}_+(\theta_1)$ and
$\mathbf{F}_+(\theta_1)$ obtained from the vectors $\mathbf{G}_{\pm}(\rho)$ through the substitution
$\rho = f \sin{\theta_1}$
\begin{subequations}
\label{Fvectros}
\begin{gather}
\mathbf{F}_+(\theta_1) = 
\big (
F_{m_1}^+(\theta_1), F_{m_2}^+(\theta_1) \ldots F_{m_p}^+(\theta_1)
\big ), \\
\mathbf{F}_- (\theta_1)=
\big (
F_{m_1}^-(\theta_1), F_{m_2}^-(\theta_1) \ldots F_{m_p}^-(\theta_1)
\big ).
\end{gather}
\end{subequations}
Below, we will call an entrance field description with such vectors followed by a respective integer-valued set $M$
a single-integral treatable representation.

Finally, one can develop a  program script to compute partial focal fields and calculate focusing of any
entrance collimated beam with use of such a script, performing the following simple steps.\\
1) {Describe the entrance beam light field in terms of amplitude vectors} $\mathbf{F}_+(\theta_1)$ {and}
$\mathbf{F}_- (\theta_1)$ {and input these vectors and a set} $M$.
Note that field of \textit{any} beam satisfying Maxwell's equations must be describable in these terms, although
in principle such a representation may be complicated and poorly treatable. \\
2) {Input microcavity mirror matrices} $M_1(\theta_1)$ {and} $M_2(\theta_1)$ {as in}
Eq.~(\ref{MirrorMatr}) {and the following other numeric parameters: the objective angular aperture
$\theta_\mathrm{max}$,
excitation wavelength $\lambda_{ex}$, core layer thickness $d$, core layer left bound $z_1$,
refractive indices $n_1$, $n$, and $n_2$.} \\
3) Obtain the resultant focal field as
\begin{align}
\label{FocFieldFinSer}
\mathbf{E}^f(\rho,\varphi,z) = 0.5 i k_1 f \mathrm{e}^{i k_1 f} \big [ E_+(\rho,\varphi,z)
\mathrm{e}^{- i \varphi} \mathbf{n}_+
\nonumber \\
+ E_-(\rho,\varphi,z) \mathrm{e}^{ i \varphi} \mathbf{n}_- + E_z(\rho,\varphi,z) \mathbf{n}_z     \big]
\end{align}
with
\begin{align}
\label{FocalAmpsFin}
E_j(\rho,\varphi,z) = \big [ \mathbf{A}_{j,+}(\rho,z) \mathrm{e}^{i (\varphi + \pi /2)} +
\nonumber \\
\mathbf{A}_{j,-}(\rho,z) \mathrm{e}^{ - i (\varphi + \pi /2)}         \big ] \cdot \mathbf{e}(\varphi + \pi /2),
\end{align}
where
\begin{equation}
\label{A_vector}
\mathbf{A}_{j,\pm}(\rho,z) = \big ( A_{j,\pm}^{m_1} (\rho,z), A_{j,\pm}^{m_2} (\rho,z)
\ldots   A_{j,\pm}^{m_p} (\rho,z)           \big )
\end{equation}
with $A_{j,s}^{m_l} (\rho,z)$ being calculated
programmatically according to Eqs.~(\ref{FieldIncComponentsIntegral}), where $j = +,-,z$, $s = +,-$, $l = [1 \ldots p]$.
Vectors $\mathbf{A}_{j,s}(\rho,z)$ are induced by vectors $\mathbf{F}_s(\theta_1)$
and both of them are $p$-dimensional vectors in contrary to spatial three-dimensional vector
$\mathbf{E}^f(\rho,\varphi,z)$ and other spatial vectors.
In step 2, a component $F_{m_j}^s(\theta_1)$ for a given entrance field $\mathbf{E}_0(\theta_1, \phi)$
can be calculated in general as
\begin{equation}
F_{m_j}^s(\theta_1) = \frac{1}{2 \pi} \int \displaylimits_{0}^{2 \pi} \mathbf{E}_0(\theta_1, \phi) \cdot \mathbf{n}_s
\exp(- i m_j \phi) \mathrm{d} \phi,
\end{equation}
although normally it is obtained in simpler ways as will be shown in the next section.
Note that calculating a dot product of complex vectors $\mathbf{E_0} \cdot \mathbf{n}_s$ one should
take complex-conjugate of the vector $\mathbf{n}_s$, according to
complex vector dot product calculation rules.
For example, if we have a vector $\mathbf{E}_0 = (5+2 i, 1 + i)$ (in Cartesian coordinates) and need to find the
dot product $\mathbf{E}_0 \cdot \mathbf{n}_+$, it will be calculated as $\mathbf{E}_0 \cdot \mathbf{n}_+ = 2^{-0.5} 
\times [ (5 + 2 i)\times \bar{1} + (1+i) \times \bar{i} ] = 2^{-0.5} \times (6+i)$.

\subsection{The infinite-series field problem}

\subsubsection{Field negligibility criterion}
If an entrance beam electric field is presented by a finite superposition of vortex beams, there is no practical issue to define
set $M$ in Eq.~(\ref{SetInteger}) and entrance vectors $\mathbf{F}_{\pm}(\theta_1)$ in Eq.~(\ref{Fvectros}),
but if it is presented by an infinite series, one needs to reveal and reject negligible terms to reduce the series
down to a finite sum.
In some cases, it might be a challenging issue and below we derive and suggest a simple practical instruction,
which allows one to reject superfluous terms.

First of all, one needs to clarify the criteria, to which a term (or group of terms) must satisfy to be considered
as negligible. Normally, such criteria are defined from an acceptable error level and we will
exploit this concept too. Let us consider a field component $E_j$ expressed by Eq.~(\ref{FocalAmpsFin})
and explore its possible calculation error. We will denote its precise value as $E_j^0$---which although exists
theoretically in general, may be neither precisely measurable, nor reliably computable in some particular cases,
but it does not prevent the present consideration---and practically computed one as $E_j^c$.
To assess the error, we should explore the difference of the two numbers $E_j^c - E_j^0$,
taking into account their complex-number nature.
There exist relative error and absolute one and we can exploit one or both of them.
The concept of the relative error encounters the potential problem of a zero-valued field magnitude:
in case $E_j = 0$ (or $E_j \to 0$), the relative error is either undefined or is of great magnitude,
which prevents its efficient use.
The concept of absolute error is free of this drawback, but to exploit it successfully, one needs to
know possible field magnitude in advance.
This is another problem, but in contrary to the zero-value one, it can be treated significantly
better (as will be shown below) and we will exploit the absolute error concept.

To work with an absolute error, we need to define units to measure the light field, for example,
volts per meter, but a field magnitude measured in volts per meter cannot be known a priori even
to the accuracy of several orders, because one can vary entrance beam power at will,
setting arbitrary value of this (at least, theoretically) to induce focal field with arbitrary absolute values.
It means that a calculated field cannot be compared
with a reference field value $E_0 = 1 \, V/m$, but it can be compared with the entrance beam field.
As a reference magnitude of an entrance beam field  its root-mean-square value 
\begin{equation}
\label{AverageEntr}
A_{en} = \frac{ \sqrt{ \int \displaylimits_{0}^{\theta_{\mathrm{max}}} \int \displaylimits_{0}^{2 \pi}
| \mathbf{E}_0(\theta_1,\phi) |^2 \sin{\theta_1} \cos{\theta_1} \mathrm{d} \phi \mathrm{d} \theta_1 } }
{\sqrt{\pi} \sin{\theta_\mathrm{max}} } 
\end{equation}
can be used with averaging over the entrance beam cross section.
To "measure" focal field in $A_{en}$, we will exploit the energy conservation concept.
Cycle-averaged electric field energy density of a field with a frequency $\omega_0$ at a location
$\mathbf{r}$ is expressed as
\begin{equation}
\label{ElEnerg}
w_E(\mathbf{r},\omega_0) = \frac{\varepsilon_0}{4} \frac{d}{d \omega} \{ \omega \mathrm{Re}
[\varepsilon (\omega, \mathbf{r}) ]  \} \Big | _{\omega = \omega_0} 
|\mathbf{E}(\mathbf{r})|^2,
\end{equation}
where $\varepsilon (\omega, \mathbf{r})$ is the permittivity \citep{novotny2012principles}.
As the field under consideration propagates along the direction $z$,  its energy inside an infinitesimal
layer of a thickness $dz$ is a constant.
In general, the derivative factor in right-hand side of Eq.~(\ref{ElEnerg}) is different in
an objective medium and the core layer, but this difference can barely exceed one or two orders
in magnitude and might be temporally neglected.
From this, the assessment
\begin{equation}
\label{FieldFocOrder}
A_f = \sqrt{ \frac{S_{en}}{S_f} } A_{en}
\end{equation}
is derived, where $A_f$ is a root-mean-square of the focal field, $S_{en}$ and $S_f$ are
the entrance beam cross-section area and focal-region target plane effective area.
The former can be expressed as $S_{en} = \pi f^2 \sin^2 {\theta_\mathrm{max}}$
and the latter can be assessed as $S_f = \pi (2 \pi \alpha / k)^2$ with $\alpha$ being a real
number of the order of magnitude from 1 to about 10.
Conceptually, the effective area is concerned to the area of a circle of radius $d = \alpha \lambda$,
inside which the most part
of the field energy is concentrated and it can be finely applied to tightly focused fields.
For example, for focused linearly polarized Gaussian beams the circle diameter normally is about from
$0.5 \lambda$ to $3 \lambda$ depending on the filling factor \citep{novotny2012principles},
for cylindrical vector beams to about $\lambda$ \citep{zhan2009cylindrical},
for vortex vector beams from $2 \lambda$ up to $6 \lambda$ \citep{zhan2006properties,
xiaoqiang2018focusing,stafeev2021tight}.
In general, an entrance beam may be poorly suitable for tight focusing and the effective radius
may be arbitrarily large, but the concept of tight focusing requires the effective circle
to be pretty small (down to the dimensions above) and we can assign the coefficient $\alpha$
to be $\alpha \le 10$.
Substituting $S_{en}$ and $S_f$ into Eq.~(\ref{FieldFocOrder}), we obtain
\begin{equation}
A_f = k f \frac{\sin{\theta_{\mathrm{max}}}}{2 \pi \alpha} A_{en}
\end{equation}
and the assessment for $\alpha$ above gives
\begin{equation}
\label{FinAssMod}
A_f \gtrsim 0.01 k f  A_{en}.
\end{equation}
From this, average values of a focal field normalized to the factor $k f A_{en}$ should amount to about $0.01$.
Additionally, the following aspects should be taken into account.
First, we neglected the derivative factor in Eq.~(\ref{ElEnerg}) and now will assume that it is
able to decrease focal field up to 1 order in magnitude.
Second, the energy conservation law was exploited, but in layered structures energy losses are
possible: let us assume that they decrease the field up to 2 orders in magnitude.
Third, one is normally interested not only in maximal and average field values, but also
in small values located in dark areas of the focal region
and we will consider field values being 2 orders of magnitude smaller than
an average one to be small and smaller values to be negligible.
Finally, a field value to be considered as a negligible component, it must be $1+2+2 = 5$ orders smaller
than an average one given by Eq.~(\ref{FinAssMod})
and for a field normalized to $k f A_{en}$ the criterion of negligibility reads as
\begin{equation}
\frac{A_f }{k f A_{en}}< 10^{-7}.
\end{equation}
For abnormal conditions---such as significant field energy losses, locations far beyond the effective
circle, extremely large refractive index of the target layer, and similar---this criterion can be
not suitable and should be properly corrected by decreasing the threshold in the right-hand side.
For normal conditions we recommend to use it with the threshold of $10^{-7}$.
This criterion can be readily applied to the field in Eq.~(\ref{FocFieldFinSer}): 
Because of the factor $k_1 f$ in the right-hand side, its components $E_j(\rho,\varphi,z)$ ($j = +,-,z$)
can be considered as fields already normalized to this factor and the negligibility criterion
for them should be reformulated as
\begin{equation}
\label{NeglCrit}
\frac{|E_j^{rest}(\rho,\varphi,z)|}{A_{en}} < 10^{-7},
\end{equation}
where a field component is divided into two parts as
\begin{equation}
E_j(\rho,\varphi,z) = E_j^{0}(\rho,\varphi,z) + E_j^{rest}(\rho,\varphi,z)
\end{equation}
with $ E_j^{0}(\rho,\varphi,z)$ being non-negligible and  $E_j^{rest}(\rho,\varphi,z)$ negligible value.

\subsubsection{Bessel-function integral series properties}
Here, we will explore the behavior of series $\mathbf{A}_{j,\pm}(\rho,z)$ expressed by Eq.~(\ref{A_vector})
to apply the suggested criterion of negligibility to them and reveal their components that can be considered
as negligible and rejected.

First, series describing entrance field amplitudes~(\ref{AmpsAsSer}) must converge and amplitudes
$F_{m_j}^s(\theta_1)$~(\ref{Fvectros}) will
decrease starting from a number $m_0^{en}$ in the positive direction of numbers (from $-\infty$ to $\infty$)
and similarly in the negative direction (the index $en$ refers to the entrance field). Below, we will investigate
only a positive direction as the series behavior is similar in both positive and negative directions and
results obtained for one of them will be valid for another.
% From this, we can suppose that starting from a number $m_0 > m_0^{en}$, focal amplitudes~(\ref{FieldIncComponentsIntegral}) will decrease and we should explore their behavior.
We will represent  focal amplitudes~(\ref{FieldIncComponentsIntegral}) in a generalized form
\begin{equation}
\label{AmplGener}
A_m(\rho,z) = \int \displaylimits_{0}^{\theta_{\mathrm{max}}} K(\theta_1)  F_m (\theta_1)
J_{l(m)}(k_1 \rho \sin{\theta_1}) T(\theta_1, z) \mathrm{d} \theta_1
\end{equation}
for the sake of simplicity.
The behavior of such a function depends on an entrance function $F_m (\theta_1)$ and Bessel function
$J_{l(m)}(k_1 \rho \sin{\theta_1})$.
From the definition of the index $m_0^{en}$ given above, the entrance field terms $F_m (\theta_1)$
do not decrease in the range $m = [0 \dots m_0^{en}]$ and we cannot guarantee their
negligibility regardless other conditions and consequently reject this region.
But at $m > m_0^{en}$ the entrance function does not prevent focal series convergence
and there exists a probability that terms starting from $m_0 > m_0^{en} $ can be omitted.
To understand it, we need to explore the behavior of Bessel function.

Bessel function of the first kind $J_n(x)$ of an integer order $n$ obeys  Bessel's differential equation
\begin{equation}
\label{BessEqClass}
x^2 \frac{\mathrm{d}^2 J_n(x)}{\mathrm{d} x^2} + x \frac{\mathrm{d} J_n(x)}{\mathrm{d} x}
+(x^2 - n^2) J_n(x) = 0
\end{equation}
and from its definition, it exhibits the following two properties:
\begin{equation}
J_{-n}(x) = (-1)^n J_n(x)
\end{equation}
and
\begin{equation}
\label{BessAtZero}
J_{n}(x) > 0, \frac{\mathrm{d} J_n(x)}{\mathrm{d} x} >0, \frac{\mathrm{d}^2 J_n(x)}{\mathrm{d} x^2} > 0
\,\,\,    \Big | _{x \to 0+}.
\end{equation}
The former guarantees that absolute value of a function behaves similarly at positive and negative
$n$'s, which allows us to extrapolate results for a positive-valued series branch to a negative-valued one;
the latter states that at small positive values of an argument $x$, Bessel function and its
first and second derivatives are positive.
Now, we will explore the function's behavior in the range $x = [0 \dots n]$ at a positive $n$.
All over this range, coefficients $x^2$ and $x$ in Eq.~\eqref{BessEqClass} are positive
and the coefficient $n^2 - x^2$ is negative.
Next, it follows from Eq.~\eqref{BessAtZero} that at small positive $x$ values, Bessel function
and its first derivative are not only positive, but also  increasing functions and 
the first and second terms in Eq.~\eqref{BessEqClass} are positive and the third one negative.
The function $J_n(x)$ to reach a local maximum and start decreasing, it is necessary its first
derivative to take on zero value and become negative.
For this, the first derivative must start decreasing preliminary, which is possible if and only if
the second derivative becomes negative.
So, at $x=0$ $J_n(x)$ and both the derivatives are equal to $0$ and increasing,
at any $x=\epsilon>0$ with $\epsilon$ being an infinitesimal value,
the function is positive and increasing and, at least, the first derivative is
positive and increasing too. From the explanation above, the second derivative
will start decreasing and take on negative values first.
At a negative second derivative and positive first one and function itself,
the first and the third terms of Eq.~\eqref{BessEqClass} will be negative and
the second one positive, which makes zero-balance possible in principle
and allows us to permit the second derivative to reach a zero-value at any
$x = x_0$ in the range $x=[0 \dots n]$ and then take on negative values.
After that, the first derivative can reach its zero value and go to the negative range, but
(1) $J_n(x)$ cannot become negative at the same $x$ value as its first derivative and
(2) the second derivative cannot become positive instantly after the first one reaches
its zero and goes to the negative range. It means that if the first derivative becomes negative,
a range $x=[x_1 \dots x_2]$ where $J_n(x)$ is positive and both its derivatives are negative
must exist inside the considered range $x = [0 \dots n]$. In such a range, all the three terms
in Eq.~\eqref{BessEqClass} will be negative, which makes zero-balance impossible
and brings us to the conclusion that in range $x=[0 \dots n]$ Bessel function $J_n(x)$
and its first derivative cannot take on negative values
\begin{equation}
\label{BessPositivity}
J_{n}(x) > 0, \frac{\mathrm{d} J_n(x)}{\mathrm{d} x} >0 \,\,\,    \Big | _{x = [ 0 \dots n]}.
\end{equation}

Next, Bessel functions obey the following universal recursive relation
\begin{equation}
J_n(x) = \frac{n+1}{x} J_{n+1}(x) + \frac{\mathrm{d} J_{n+1}(x)}{\mathrm{d} x}.
\end{equation}
From this and Eq.~\eqref{BessPositivity}, it directly follows that in range $x = [0 \dots n]$ an inequality
\begin{equation}
\label{bess_enequ}
J_{n+p+1}(x) < \frac{n}{n +p +1} J_{n+p}(x) 
\end{equation}
for $p = [0, 1, 2 \dots \infty]$ is valid.
Applying this to Eq.~(\ref{AmplGener}), we conclude that the Bessel function decreases with an increasing number
at $m > m_0^{bf} $, where $m_0^{bf}$ is defined from the equation
\begin{equation}
\label{BesselIndex354}
l \big ( m_0^{bf} \big ) = [k \rho \sin{\theta_\mathrm{max}}]+1
\end{equation}
and $[\alpha]$ denotes integer part of a real number $\alpha$.
So, starting from $m_0 = \sup \{m_0^{bf}, m_0^{en} \} $ ($m_0^{en}$ was defined above as a number
from which the entrance coefficient starts decrease), both functions $F_m(\theta_1)$ and $J_{l(m)}(k \rho
\sin{\theta_1})$ decrease with an increasing number.

A residual part of a partial amplitude in Eq.~(\ref{FocalAmpsFin}) takes the form
\begin{equation}
E^{\pm}_{j,rest}(\rho,\varphi,z) = \mathrm{e}^{\pm (\varphi + \pi /2)} \sum_{k=m_0+p+1}^{\infty}
A^k_{j,\pm}(\rho,z) \mathrm{e} ^ {i k (\varphi + \pi /2)}
\end{equation}
and we need to satisfy the condition of its negligibility
\begin{equation}
|E^{\pm}_{j,rest}(\rho,\varphi,z)| < \varepsilon,
\end{equation}
where $\varepsilon$ is an acceptable error level.
From inequality~(\ref{bess_enequ}), we have
\begin{equation}
 \sum_{k=l(m_0)+p+1}^{\infty} J_{k} (x) < \Bigg (1 + \frac{l(m_0)}{p+1} \Bigg) J_{ l(m_0) + p} (x)
\end{equation}
and
\begin{align}
|E^{\pm}_{j,rest}(\rho,\varphi,z)| < \Bigg (1 + \frac{l(m_0)}{p+1} \Bigg )
\nonumber \\
\times \int \displaylimits_{0}^{\theta_\mathbf{max}}
| K(\theta_1)  F_{m_0 + p} (\theta_1) | J_{ l(m_0) + p} (k_1 \rho \sin{\theta_1}) \mathrm{d} \theta_1.
\end{align}
Thus, to define index $p$ and set entrance vectors $\mathbf{E}_{\pm}$ given by Eq.~(\ref{Fvectros}),
we should asign $p$ the least value at which the condition
\begin{align}
\label{FieldRest}
\Bigg [ 1 + \frac{l(m_0)}{p+1} \Bigg ] \int \displaylimits_{0}^{\theta_\mathbf{max}}
| K(\theta_1)  F_{m_0 + p} (\theta_1) |
\nonumber \\
\times J_{ l(m_0) + p} (k_1 \rho \sin{\theta_1}) \mathrm{d} \theta_1 < \varepsilon
\end{align}
is satisfied. Applying criterion~\eqref{NeglCrit}, we obtain the final equation for index $p$
\begin{align}
\label{FieldRest_det}
\Bigg [ 1 + \frac{l(m_0)}{p+1} \Bigg ] \int \displaylimits_{0}^{\theta_{max}}
\frac{| K(\theta_1)  F_{m_0 + p} (\theta_1) |}{A_{en}}
\nonumber \\
\times J_{ l(m_0) + p} (k \rho \sin{\theta}) \mathrm{d} \theta_1 < 10^{-7}
\end{align}
and the positive branch of the series will be described by a set $M_{pos} = \{0,1,2 \dots m_0+p \}$.
The negative branch is obtained in a similar way.

\section{Examples of practical calculations and enhancement of the single-integral algorithm}
\label{Computations4527}

In this section, we will consider some examples of possible practical implementation of the suggested algorithm
to illustrate its applicability and disclose some details.
First, several widely used azimuth-continued entrance laser fields will be considered and 
represented in the single-integral treatable form.
Next, a fractional vortex beam will be considered and analyzed numerically as an example of an
azimuth-discontinued beam.
Finally, the suggested algorithm will be tested as a tool for calculations of focal fields in finite spacial
regions and some details of the algorithm will be corrected to adapt it for this task.

\subsection{Finite-series fields: Examples of single-integral treatable representations}

Let us consider a linearly polarized beam with a vector amplitude $\mathbf{E}_0 (\theta_1, \phi) = 
P(\theta_1) \mathbf{n}_x$, where $P(\theta_1)$ is a scalar amplitude.
Depending on the scalar amplitude, it can be a true plane wave beam, Gaussian beam,
Hermite-Gaussian mode, Laguerre-Gaussian mode,
and any other. For example, for a Gaussian paraxial beam, we have the amplitude
$P(\theta_1) = C \exp [ - \sin^2{\theta_1} / (f_0 \sin^2{\theta_\mathrm{max}})]$, where $C$ is a constant,
$f_0$ the filling factor, $\theta_\mathrm{max}$ the objective angular aperture \citep{novotny2012principles}.
To represent such a field in a single-integral treatable form, we need to express the vector $\mathbf{n}_x$
through circular polarization vectors. From Eq.~(\ref{Circ_in_Cart}), one obtains $\mathbf{n}_x = 2^{-0.5}
(\mathbf{n}_+ + \mathbf{n}_-)$ and we can write the respective single-integral treatable representation
straightforward as follows: $M=\{0\}$, $\mathbf{F}_+(\theta_1) = (2^{-0.5} P(\theta_1))$, and
$\mathbf{F}_-(\theta_1) = (2^{-0.5} P(\theta_1))$. Here, $M$ is a single-component set and
$\mathbf{F}_{\pm}(\theta_1) $ are single-component vectors.

The next entrance beam that we will analyze is an elliptically polarized cylindrical vector Bessel beam
\citep{boichenko2018theoretical, boichenko2020toward}. Mathematically it is expressed as
$\mathbf{E}_0(\theta_1, \phi) = J_1(\beta \sin{\theta_1}) ( \mathrm{e}^{i \Phi} \cos{\alpha} \, 
\mathbf{n}_{\rho} +
\mathrm{e}^{- i \Phi} \sin{\alpha} \, \mathbf{n}_{\varphi} )$ with $\beta$ being a constant,
$\alpha$ and $\Phi$ ellipticity parameters, $J_1(x)$ Bessel function of the first kind of a variable $x$.
Again, we need to represent it in terms of circular polarization vectors.
For this, we exploit Eqs.~(\ref{CVBtoCircPol}) and obtain $M=\{ -1, 1 \}$,
$\mathbf{F}_+(\theta_1) = ( 2^{-0.5} J_1(\beta \theta_1) ( \mathrm{e}^{i \Phi} \cos{\alpha} - i
\mathrm{e}^{- i \Phi} \sin{\alpha}  ) , 0)$,
$\mathbf{F}_-(\theta_1) = ( 0, 2^{-0.5} J_1(\beta \theta_1) ( \mathrm{e}^{i \Phi} \cos{\alpha}  + i
\mathrm{e}^{- i \Phi} \sin{\alpha}  ) )$.

In \citep{kotlyar2021transverse} tight focusing of laser beams with hybrid circular-azimuthal polarization
was theoretically studied. Mathematically, such a light field is expressed as $\mathbf{E}_0(\theta_1,\phi) =
A(\theta_1) [ -( i \sin m \phi) \mathbf{n}_x + (\cos m \phi) \mathbf{n}_y]$, where $A(\theta_1)$ is a scalar
amplitude and $m$ an integer constant. It can be represented in single-integral treatable form as well as the
fields considered above. Exploiting Euler's formula to express the trigonometric functions in terms of complex
exponents, we obtain $M = \{ m, -m \}$, $\mathbf{F}_+(\theta_1) = -0.5 i A(\theta_1) (\mathrm{e}^{- i \pi /4},
\mathrm{e}^{i \pi /4})$, and $\mathbf{F}_-(\theta_1) = 0.5 i A(\theta_1) (\mathrm{e}^{ i \pi /4},
\mathrm{e}^{-i \pi /4})$.

\subsection{Logarithm-like series entrance field: Common investigation}
\label{LogSeries1425}

So, we considered several azimuth-continued fields as examples and now its time to explore a field with a fractional
topological charge, which is an example of an azimuth-discontinued field.
An extensive review of fractional vortex beams was performed, for example, in \citep{zhang2021review}.
Such a beam can be described by any scalar amplitude but the simplest possible one is a fractional Bessel beam
and we will consider the following one
\begin{equation}
\label{FracBess5248}
\mathbf{E}_0(\theta_1,\phi) = J_{\alpha}(2 \sin{\theta_1}) \mathrm{e}^{i \alpha \phi} \mathbf{n}_+
\end{equation}
with $J_{\alpha}(x)$ being Bessel function of the first kind and $\alpha$ a non-integer positive number.
Its scalar amplitude is decomposed as \citep{berry2004optical}
\begin{equation}
\label{SeriesLog}
E_0^+(\theta_1,\phi) = \frac{\exp(i \pi \alpha) \sin(\pi \alpha)}{\pi} J_{\alpha}(2 \sin{\theta_1})
\sum_{m=-\infty}^{\infty}  \frac{\mathrm{e}^{i m \phi}}{\alpha - m}
\end{equation}
and $M = \{-\infty \ldots -1, 0, 1 \ldots \infty \}$, $\mathbf{F}_+(\theta_1) = [ \exp(i \pi \alpha) \sin(\pi \alpha)/ \pi]
J_{\alpha}(2 \sin{\theta_1}) \big (\ldots \frac{1}{\alpha + 1}, \frac{1}{\alpha}, \frac{1}{\alpha - 1} \ldots \big )$
with $F_m^+(\theta_1) =   [ \exp(i \pi \alpha) \sin(\pi \alpha)/ \pi] J_{\alpha}(2 \sin{\theta_1}) / ( \alpha - m)$.

Here, we will investigate the focusing of a beam with $\alpha = 0.4$ by an objective with 
an angular aperture $\theta_{\max} = 64^\circ$ for the following two cases:
(1) free space with the refractive index $n = 1.5$ and an excitation beam with the wavelength
$\lambda = 0.5$~$\mu m$,
(2) a planar microcavity exploited in \citep{rammler2022strong, rammler2023analysis} with
mirrors described by the matrices
\begin{widetext}
\begin{subequations}
\label{MirrorMatrMeixner}
\begin{gather}
M_1 =
\begin{pmatrix}
3 \, nm & 30 \, nm & 5 \, nm & 20 \, nm \\
1.47 & 1.5159+1.8844 i &	0.10433+2.5279 i & 2.2288+4.3757 i
\end{pmatrix}, \label{Mirror1M_example} \\
 \label{Mirror2M_example}
M_2 = 
\begin{pmatrix}
3 \, nm & 60 \, nm & 5 \, nm & 20 \, nm \\
1.47 & 1.5159+1.8844 i&	0.10433+2.5279 i & 2.2288+4.3757i
\end{pmatrix},
\end{gather}
\end{subequations}
\end{widetext}
the laser excitation wavelength $\lambda = 0.44$~$\mu m$, core layer refractive index
$n = 1.34$ (water) and thickness $d = 1$~$\mu m$, surrounding media refractive indices
$n_1 = 1.526$ and $n_2 = 1.526$
(glass), core layer edge position $z_1 = 0.3$~$\mu m$.
The mirrors described by Eqs.~(\ref{MirrorMatrMeixner}) are assumed to consist
of the following four layers: silicon dioxide $SiO_2$, gold, silver, and chromium.
To calculate partial fields~\eqref{FieldIncComponentsIntegral} normalized to 
the average entrance field amplitude~\eqref{AverageEntr} and form
vectors~\eqref{A_vector}, we developed a script in Python 3.9.
Another script was developed to compute normalized focal field according to the
double-integral procedure for comparing.

First, we analyzed the behavior of the residual field described by the left-hand side of
Eq.~(\ref{FieldRest}) in free space. The respective results at the distance coordinate
$\rho = 3 \lambda = 1.5$~$\mu m$ and longitudinal coordinate $z=0$ are displayed in
Table~\ref{FS_rhoBig}. The positive-valued branch is considered; negative-valued one
behaves similarly and we do not show it.
\begin{table}[h]
\caption{\label{FS_rhoBig}
The positive-branch residual field as a function of a series number in free space. The parameters are
assumed to be as follows: $\lambda = 0.5$~$\mu m$, $n = 1.5$, $z = 0$, $\rho = 3 \lambda$.
}
\begin{ruledtabular}
\begin{tabular}{cccc}
$m$ & $\max(|E_+|_{res})$ & $\max(|E_-|_{res})$ & $\max(|E_z|_{res})$ \\
\hline
26	&	$	4.41 \times 10^{-03	}	$	&	$	4.57 \times 10^{-04	}	$	&	$	2.04 \times 10^{-03	}	$	\\
27	&	$	1.29 \times 10^{-03	}	$	&	$	1.21 \times 10^{-04	}	$	&	$	5.69 \times 10^{-04	}	$	\\
28	&	$	4.75 \times 10^{-04	}	$	&	$	4.08 \times 10^{-05	}	$	&	$	2.00 \times 10^{-04	}	$	\\
29	&	$	1.87 \times 10^{-04	}	$	&	$	1.47 \times 10^{-05	}	$	&	$	7.52 \times 10^{-05	}	$	\\
30	&	$	7.45 \times 10^{-05	}	$	&	$	5.40 \times 10^{-06	}	$	&	$	2.88 \times 10^{-05	}	$	\\
31	&	$	2.96 \times 10^{-05	}	$	&	$	1.98 \times 10^{-06	}	$	&	$	1.10 \times 10^{-05	}	$	\\
32	&	$	1.16 \times 10^{-05	}	$	&	$	7.18 \times 10^{-07	}	$	&	$	4.13 \times 10^{-06	}	$	\\
33	&	$	4.42 \times 10^{-06	}	$	&	$	2.56 \times 10^{-07	}	$	&	$	1.52 \times 10^{-06	}	$	\\
34	&	$	1.65 \times 10^{-06	}	$	&	$	8.94 \times 10^{-08	}	$	&	$	5.50 \times 10^{-07	}	$	\\
35	&	$	6.02 \times 10^{-07	}	$	&	$	3.05 \times 10^{-08	}	$	&	$	1.94 \times 10^{-07	}	$	\\
36	&	$	2.14 \times 10^{-07	}	$	&	$	1.02 \times 10^{-08	}	$	&	$	6.68 \times 10^{-08	}	$	\\
37	&	$	7.39 \times 10^{-08	}	$	&	$	3.31 \times 10^{-09	}	$	&	$	2.24 \times 10^{-08	}	$	\\
38	&	$	2.49 \times 10^{-08	}	$	&	$	1.05 \times 10^{-09	}	$	&	$	7.34 \times 10^{-09	}	$	\\
39	&	$	8.19 \times 10^{-09	}	$	&	$	3.27 \times 10^{-10	}	$	&	$	2.34 \times 10^{-09	}	$	\\
40	&	$	2.63 \times 10^{-09	}	$	&	$	9.93 \times 10^{-11	}	$	&	$	7.30 \times 10^{-10	}	$	\\
41	&	$	8.20 \times 10^{-10	}	$	&	$	2.94 \times 10^{-11	}	$	&	$	2.22 \times 10^{-10	}	$	\\
42	&	$	2.50 \times 10^{-10	}	$	&	$	8.53 \times 10^{-12	}	$	&	$	6.60 \times 10^{-11	}	$	\\
43	&	$	7.45 \times 10^{-11	}	$	&	$	2.41 \times 10^{-12	}	$	&	$	1.92 \times 10^{-11	}	$	\\
44	&	$	2.17 \times 10^{-11	}	$	&	$	6.69 \times 10^{-13	}	$	&	$	5.44 \times 10^{-12	}	$	\\
45	&	$	6.15 \times 10^{-12	}	$	&	$	1.81 \times 10^{-13	}	$	&	$	1.51 \times 10^{-12	}	$	\\
46	&	$	1.71 \times 10^{-12	}	$	&	$	4.81 \times 10^{-14	}	$	&	$	4.09 \times 10^{-13	}	$	\\
\end{tabular}
\end{ruledtabular}
\end{table}
The first thing we need to do to explore the function behavior is defining the critical index $m_0$.
The entrance field  is described by a monotonically decreasing series~\eqref{SeriesLog} and
the entrance-function decrease beginning index $m_0^{en}$ is not important in the present case,
which means that $m_0 = m_0^{bf}$ and the critical index should be calculated from
Eq.~\eqref{BesselIndex354} as
\begin{equation}
l(m_0) = \Bigg [\frac{2 \pi n_1}{\lambda} 3 \lambda \sin 64^\circ \Bigg ] + 1 = 26.
\end{equation}
The index $l(m_0)$ is expressed in terms of $m_0$ as $l = m_0$ for $A_+$ component,
$l=m_0+2$ for $A_-$ component, and $l=m_0+1$ for $A_z$ component, which can be obtained
from comparing Eq.~\eqref{AmplGener} with \eqref{FieldIncComponentsIntegral}.
So, $m_0 = 26$ for the plus-component, 24 for the minus one, and 25 for the longitudinal one.
To put it strictly, it would be preferable to start indices in the table from the respective values
for each component, but on the one hand we can start after but not before the critical index
and on the other hand the differences of 1 and 2 steps might be considered as negligible
and we start from $m=26$.
Table~\ref{MC_rhoBig} displays similar results for the microcavity described above.
\begin{table}[h]
\caption{\label{MC_rhoBig}
The positive-branch residual field as a function of a series number in a metal-dielectric planar microcavity.
The cavity parameters are described in the text, coordinates are $\rho = 3 \lambda$, $z=0$.
}
\begin{ruledtabular}
\begin{tabular}{cccc}
$m$ & $\max(|E_+|_{res})$ & $\max(|E_-|_{res})$ & $\max(|E_z|_{res})$ \\
\hline
26	&	$	3.84 \times 10^{-04	}	$	&	$	1.24 \times 10^{-04	}	$	&	$	4.24 \times 10^{-04	}	$	\\
27	&	$	1.12 \times 10^{-04	}	$	&	$	3.28 \times 10^{-05	}	$	&	$	1.19 \times 10^{-04	}	$	\\
28	&	$	4.11 \times 10^{-05	}	$	&	$	1.10 \times 10^{-05	}	$	&	$	4.24 \times 10^{-05	}	$	\\
29	&	$	1.61 \times 10^{-05	}	$	&	$	3.97 \times 10^{-06	}	$	&	$	1.61 \times 10^{-05	}	$	\\
30	&	$	6.42 \times 10^{-06	}	$	&	$	1.46 \times 10^{-06	}	$	&	$	6.20 \times 10^{-06	}	$	\\
31	&	$	2.54 \times 10^{-06	}	$	&	$	5.38 \times 10^{-07	}	$	&	$	2.38 \times 10^{-06	}	$	\\
32	&	$	9.90 \times 10^{-07	}	$	&	$	1.95 \times 10^{-07	}	$	&	$	9.04 \times 10^{-07	}	$	\\
33	&	$	3.78 \times 10^{-07	}	$	&	$	6.98 \times 10^{-08	}	$	&	$	3.36 \times 10^{-07	}	$	\\
34	&	$	1.41 \times 10^{-07	}	$	&	$	2.44 \times 10^{-08	}	$	&	$	1.22 \times 10^{-07	}	$	\\
35	&	$	5.13 \times 10^{-08	}	$	&	$	8.33 \times 10^{-09	}	$	&	$	4.34 \times 10^{-08	}	$	\\
36	&	$	1.82 \times 10^{-08	}	$	&	$	2.79 \times 10^{-09	}	$	&	$	1.50 \times 10^{-08	}	$	\\
37	&	$	6.27 \times 10^{-09	}	$	&	$	7.58 \times 10^{-10	}	$	&	$	5.07 \times 10^{-09	}	$	\\
38	&	$	2.10 \times 10^{-09	}	$	&	$	3.75 \times 10^{-10	}	$	&	$	1.67 \times 10^{-09	}	$	\\
39	&	$	5.56 \times 10^{-10	}	$	&	$	1.15 \times 10^{-10	}	$	&	$	4.62 \times 10^{-10	}	$	\\
40	&	$	2.85 \times 10^{-10	}	$	&	$	3.48 \times 10^{-11	}	$	&	$	1.44 \times 10^{-10	}	$	\\
41	&	$	8.81 \times 10^{-11	}	$	&	$	1.02 \times 10^{-11	}	$	&	$	4.41 \times 10^{-11	}	$	\\
42	&	$	2.66 \times 10^{-11	}	$	&	$	2.94 \times 10^{-12	}	$	&	$	1.32 \times 10^{-11	}	$	\\
43	&	$	7.84 \times 10^{-12	}	$	&	$	8.27 \times 10^{-13	}	$	&	$	3.84 \times 10^{-12	}	$	\\
44	&	$	2.25 \times 10^{-12	}	$	&	$	2.27 \times 10^{-13	}	$	&	$	1.09 \times 10^{-12	}	$	\\
45	&	$	6.34 \times 10^{-13	}	$	&	$	6.12 \times 10^{-14	}	$	&	$	3.05 \times 10^{-13	}	$	\\
46	&	$	1.74 \times 10^{-13	}	$	&	$	1.61 \times 10^{-14	}	$	&	$	8.34 \times 10^{-14	}	$	\\
\end{tabular}
\end{ruledtabular}
\end{table}
We can see that according to the negligibility criterion~\eqref{FieldRest_det},
terms starting from $m=37$  in free space and from $m = 35$ in the microcavity can be rejected.
In case one needs to calculate the field with a higher accuracy, more terms should be taken
and we can see that the series converges rapidly: to increase the accuracy by one order,
two additional terms are normally required.
Note that the entrance series expressed by Eq.~\eqref{SeriesLog} is a slowly (and conditionally)
convergent logarithmic-like series and can be considered as an example of a series
with the slowest possible convergence.
From this, we conclude that focusing process transforms any entrance series 
into a rapidly convergent one, and focal field can be calculated using pretty
small number of terms (from several terms up to several tens of them).
The only possible exception to this rule is an entrance series with rapidly increasing
coefficients at partial vortex beams in a wide index (topological charge) range, 
but such an entrance beam would be an exceptional (although, theoretically possible) example
and it is barely exploited in practice.

Further, we calculated the field inside the microcavity at several spatial points, using the single-integral
series algorithm and conventional double-integral algorithm to compare field values and computation times.
The respective results are presented in Table~\ref{MC_rhoVary}.
\begin{table}[h]
\caption{\label{MC_rhoVary}
A normalized light field inside the metal-dielectric microcavity. The cavity parameters are described in the text,
the distance coordinate is varied (column 1), $z=0$, times $t_1$ and $t_2$ correspond to the single- and
double-integral algorithms, respectively; $\delta \mathbf{E}$ is the difference between field
values calculated using the single- and double-integral algorithms.
}
\begin{ruledtabular}
\begin{tabular}{ccccc}
$\rho/\lambda$ & $M$, $\mathbf{E}$ & $t_{2}$ (s) & $t_{1}$ (s) & $\delta \mathbf{E}$ \\
\hline
\rule{0pt}{5ex}
% First line
0 &
\shortstack{$\{-2,-1,0 \}$, \\
$\begin{pmatrix} %Field
0.011-0.034 i \\
 (1.5+0.8 i) \times 10^{-3}\\
(-4.9+6.2 i) \times 10^{-3}
\end{pmatrix}$}
&
$\begin{pmatrix} %t2
12 \\
11 \\
5
\end{pmatrix}$
&
$\begin{pmatrix} %t1
0.7 \\
0.6\\
0.3
\end{pmatrix}$
&
$\begin{pmatrix} % error
3 \times 10^{-9}\\
5 \times 10^{-10} \\
2 \times 10^{-13}
\end{pmatrix}$ \\
%End first line
%-----------------------------------------------
\rule{0pt}{5ex}
% Second line
0.5 &
\shortstack{ \{ -13 \ldots 11 \}, \\
$\begin{pmatrix} %Field
(-3.8-3.0 i) \times 10^{-3} \\
(-6.1-0.5 i) \times 10^{-3}\\
(-8.9 - 2.2 i) \times 10^{-3}
\end{pmatrix}$ }
&
$\begin{pmatrix} %t2
83 \\
58 \\
27
\end{pmatrix}$
&
$\begin{pmatrix} %t1
13 \\
13 \\
5
\end{pmatrix}$
&
$\begin{pmatrix} % error
1 \times 10^{-9}\\
8 \times 10^{-10} \\
8 \times 10^{-11}
\end{pmatrix}$  \\
%End second line
%---------------------------------------------------
\rule{0pt}{5ex}
% 3rd line
1 &
\shortstack{ \{ -18 \ldots 17 \}, \\
$\begin{pmatrix} %Field
(-2.8-7.6 i) \times 10^{-4} \\
(1.7+0.4 i) \times 10^{-3}\\
(7.8 - 7.5 i) \times 10^{-4}
\end{pmatrix}$ }
&
$\begin{pmatrix} %t2
84 \\
81 \\
33
\end{pmatrix}$
&
$\begin{pmatrix} %t1
19 \\
19 \\
7
\end{pmatrix}$
&
$\begin{pmatrix} % error
4 \times 10^{-10}\\
2 \times 10^{-9} \\
1 \times 10^{-9}
\end{pmatrix}$  \\
%End 3rd line
%-------------------------------------------------------
\rule{0pt}{5ex}
% 4th line
2 &
\shortstack{ \{ -27 \ldots 27 \}, \\
$\begin{pmatrix} %Field
(-5.5-0.1 i) \times 10^{-4} \\
(-2.2 - 3.5 i) \times 10^{-4}\\
(-1.4+0.3 i) \times 10^{-3}
\end{pmatrix}$ }
&
$\begin{pmatrix} %t2
179 \\
173 \\
72
\end{pmatrix}$
&
$\begin{pmatrix} %t1
29 \\
29 \\
12
\end{pmatrix}$
&
$\begin{pmatrix} % error
9 \times 10^{-11}\\
4 \times 10^{-9} \\
4 \times 10^{-9}
\end{pmatrix}$  \\
%End 4th line
%---------------------------------------------------------
\rule{0pt}{5ex}
% 5th line
3 &
\shortstack{ \{ -36 \ldots 37 \}, \\
$\begin{pmatrix} %Field
(-1.0 + 8.8 i) \times 10^{-4} \\
(-9.8-1.9 i) \times 10^{-4}\\
(8.8+6.7 i) \times 10^{-4}
\end{pmatrix}$ }
&
$\begin{pmatrix} %t2
202 \\
197 \\
90
\end{pmatrix}$
&
$\begin{pmatrix} %t1
40 \\
39 \\
16
\end{pmatrix}$
&
$\begin{pmatrix} % error
5 \times 10^{-10}\\
5 \times 10^{-9} \\
5 \times 10^{-9}
\end{pmatrix}$  \\
%End 5th line
%---------------------------------------------------------
\end{tabular}
\end{ruledtabular}
\end{table}
The first column of the table displays the distance coordinate measured in entrance beam wavelengths.
The second one presents the set of terms $M$ and calculated entrance-normalized field values as a vector
$\mathbf{E} = (E_+,E_-,E_z)$.
The distance coordinate $\rho$ is varied as shown in the table, the longitudinal coordinate is $z = 0$,
and the azimuthal angle $\varphi = 0$.
The third and fourth columns display computation times for the single- and double-integral algorithms, respectively.
The last column shows the difference vector $\delta \mathbf{E} = (\delta E_+, \delta E_-,\delta E_z)$
with
\begin{equation}
\delta E_j = \Big | E_j^{(1)} -  E_j^{(2)} \Big |,
\end{equation}
where $j =+,-,z$ and $E_j^{(1)}$ and $E_j^{(2)}$ are field values calculated using
single- and double-integral algorithm, respectively.
The former was accomplished through the Scipy function \textit{integrate.quad}  with the
absolute error ($epsabs$) set to be equal to $10^{-9}$ and the relative error ($epsrel$) $10^{-5}$,
the latter through \textit{integrate.dblquad} with $epsabs = 10^{-7}$ and $epsrel = 10^{-4}$;
other parameters of both functions took on default values.
Assuming a field value calculated from the double-integral procedure to be a reference value
and using it to assess the error of the single-integral calculation,
we can state that the error is less than both the absolute threshold $10^{-7}$ 
and the product of the relative error (equal to $10^{-4}$) and a field absolute value
$10^{-8}$, which means that the single-integral algorithm
provides completely acceptable precision.
The comparison of times $t_1$ and $t_2$ for each field component shows that single-integral
algorithm normally computes the field several times faster than the double-integral one.
To compare these times more precisely, we collected them in Table~\ref{MC_time}.
\begin{table}[h]
\caption{\label{MC_time}
Computation times required for a total field, extracted from Table~\ref{MC_rhoVary}.
}
\begin{ruledtabular}
\begin{tabular}{cccc}
$\rho/\lambda$ & $t_{2}$ (s) & $t_{1}$ (s) & $t_2 / t_1$ \\
\hline
% First line
0 & 28 & 1.6 & 17.5 \\
%End first line
%-----------------------------------------------
% Second line
0.5 & 168 & 31 & 5.4 \\
%End second line
%---------------------------------------------------
% 3rd line
1 & 198 & 45 & 4.4\\
%End 3rd line
%-------------------------------------------------------
% 4th line
2 & 424 & 70 & 6.1\\
%End 4th line
%---------------------------------------------------------
% 5th line
3 & 489 & 95 & 5.1 \\
%End 5th line
%---------------------------------------------------------
\end{tabular}
\end{ruledtabular}
\end{table}
This table shows times required to calculate the total three-component field.
For example, Table~\ref{MC_rhoVary} shows that to compute the field through the double-integral
procedure at $\rho = 0$, one needs 12 seconds for $E_+$, 11 seconds for $E_-$, and 5
seconds for $E_z$. Table~\ref{MC_time} collects it in the first row of the second column as
$12+11+5=28$ seconds to calculate the total field.
The fourth column shows that double-integral time is 17.5 times greater than single-integral one
at $\rho = 0$, and about 5 times at $\rho$ from $0.5 \lambda$ to $3 \lambda$.
Averaged over this coordinate range ($\rho = 0 \dots 3 \lambda$) times $t_2$ and $t_1$
are $\overline{t_2} = 303 \, s$ and $\overline{t_1} = 56 \, s$ and their relation is
$\overline{t_2}/ \overline{t_1} = 5.4$.

\subsection{Computation of focal fields in finite spacial regions}

\subsubsection{Cylindrical-coordinate scanning}

The procedure of computing a focal field in a given spatial region (for example, in focal plane)
is routinely exploited today and can be considered as one of the most important tasks,
where focal field calculation is accomplished,
and here we will discuss the problem of computational time in view of this task
to compare computational times in the single- and double-integral algorithms.
Let us consider a case, when we want to calculate a field component $E_j(x,y)$ inside a square
with $x = [-2 \lambda \dots 2\lambda]$ and $y = [-2 \lambda \dots 2\lambda]$.
To perform such a calculation, one needs to define scanning steps $\Delta x$ and $\Delta y$
and the respective numbers of steps.
At a number of steps $N = 100$, scanning steps $\Delta x$ and $\Delta y$ will amount to
$[2 \lambda - (-2 \lambda)]/100 = 0.04 \lambda$. At $\lambda = 500$~$nm$, the steps
will amount to $20$~$nm$ and such values of scanning steps, region of interest dimensions,
and numbers of steps can be considered as typical.
As we need to scan the area in both $x$ and $y$ coordinates, the total number of steps
will be $N \times N = N^2$.
To calculate a field in a point with coordinates $(x,y)$, using double-integral algorithm,
we need to express these coordinates through the respective polar coordinates $(\rho,\varphi)$, 
then input these polar coordinates in Eq.~\eqref{FieldIntegral} and compute the double integral.
(It is not necessary to express Cartesian coordinates in terms of polar ones, but
we do this for the sake of obviousness.)
In the present example the longitudinal coordinate $z$ is fixed and set to be zero,
but focal region is normally  scanned plane-by-plane in $z$-direction and calculative times obtained
for a single plane can be directly extrapolated to a three-dimensional region.
Here, we have to focus one's attention at the fact that in double-integral procedure a focal field is not factorized
into two parts, one of which depends only on $\rho$ and $z$ and the other one only on $\varphi$,
and we need to calculate it in each point $(\rho,\varphi)$ separately, performing $N^2$ steps.
In contrary, in single-integral algorithm a target field is factorized into two $p$-dimensional vectors
$\mathbf{G}_\pm (\rho)$ [Eq.~\eqref{VectorAmpl746}] and $\mathbf{e}(\varphi)$ [Eq.~\eqref{ExpVector74}]
and, in fact, the only time-consuming procedure is calculating $\rho$-dependent vectors,
because both calculation of $p$ exponents in Eq.~\eqref{ExpVector74} and
multiplication of vectors $\mathbf{G}_\pm (\rho)$ and $\mathbf{e}(\varphi)$ in Eq.~\eqref{AmpsAsSer}
consume absolutely negligible time compared to integration procedures.
It provides a possibility to calculate vectors $\mathbf{G}_\pm(\rho)$ once and exploit them
at all points of the region of interest and below we will study it.

Computing a field inside a given region of a focal plane, one can exploit one of the following two
different scan-region approaches: (1) take a square region of interest (ROI) and scan it line-by-line
with lines along $x$-direction ($xy$-scan approach), (2) take a ROI of any---not necessarily
square---form and scan it in a way different from line-by-line $xy$-scan (curvilinear-scan approach).
The former may be the only possible in some practical cases: for example, if one wants to simulate
a laser-scanning fluorescence confocal image to compare it with an experimentally recorded one,
they need to reproduce the device ROI geometry, which is normally a line-by-line scanned square;
the latter can simplify computations, where it can be applied.
For example, the focal-plane square considered in the previous paragraph can be treated as follows.
We have a square with $x=[-2\lambda \dots 2\lambda]$ and $\Delta x = 0.04 \lambda$,
$y=[-2\lambda \dots 2\lambda]$ and $\Delta y = 0.04 \lambda$, $N=100$ steps in each direction,
$N \times N = N^2 = 10^4$ points.
Such an area can be covered by a circle of a radii $L/\sqrt{2} =2 \sqrt{2} \lambda$
with $L = 2 \lambda - (- 2 \lambda) = 4 \lambda$ being the square side
and this circle can readily be scanned in cylindrical coordinates instead of Cartesian ones.
We can scan it, for example, circle-by-circle with the step between circles $\Delta \rho =\Delta x= 0.04 \lambda$
($\rho =\rho_0, \rho_1, \rho_2 \dots \rho_l \dots= 0, \Delta \rho, 2 \Delta \rho, \dots l \Delta \rho \dots$
is a radii of a current circle)
and $6 \rho_l / \Delta \rho = 6 l$ points on a circle with a radii $\rho_l$ (except for the center as
at $l=0$, we have one point) and such a scan allows one
to keep distances between a given point and each of its four nearest neighbor points
equal to $\Delta \rho$.
The number of circles will amount to $M = (L / \sqrt{2}) / \Delta \rho \to [N/\sqrt{2}] = 70$,
where $[\dots]$ means an integer part.
The total number of points inside the circle amounts to $3 M (M+1)$ and inside the target square
it is assessed as $3 M (M+1) L^2 / [\pi (L / \sqrt{2})^2 ] \approx (3/\pi) N^2 \approx 0.95 N^2$,
which is $5 \%$ less than that for $xy$-scan and can be considered as acceptable.
Now, calculative times for the single- and double-integral algorithms can be assessed
and we will perform this assessment for the entrance beam~\eqref{FracBess5248}.
Taking the number of points equal to $0.95 \times N^2 = 0.95 \times 10^2 = 9500$
and one-value average times $\overline{t_2} = 303 \, s$ and $\overline{t_1} = 56 \, s$
as was obtained in the previous subsection,
one obtains $T_2 = \overline{t_2} \times 9500 = 303 \, s \times 9500 \approx 800 \, h$
for the double-integral algorithm and $T_1 = \overline{t_1} M = 56 \, s \times 70 \approx 1.1 \, h$
for the single-integral algorithm.
This assessment does not take into account computing the exponential vector $\mathbf{e}(\varphi + \pi/2)$
and scalar product  in Eq.~\eqref{FocalAmpsFin}, but they are negligible compared to $1 \, s$
and can be readily neglected. Thus, we conclude that for the entrance field under investigation,
the single-integral algorithm is 3-order faster than the
double-integral one for computations of focal fields in spacial regions where a cylindrical scan is acceptable.

In the next step, we will perform the same calculation time analysis for some pure circularly polarized
vector vortex Bessel entrance beams. An entrance beam light field is expressed as
\begin{equation}
\mathbf{E}_0(\theta,\phi) = J_m(2 \sin{\theta}) \mathrm{e}^{ i m \phi} \mathbf{n}_+
\end{equation}
and the medium is free space as described in Subsection~\ref{LogSeries1425}.
These beams are basic elements of any entrance beam Fourier series representation that
we exploit and their computation times are, in fact, the least possible.
The results for calculative times are collected in Table~\ref{FS_time_m_vary},
where $m$ is the topological charge, $t_1$ and $t_2$ are average calculative times over the region
$\rho = [0 \dots 3 \lambda]$ with the step of $0.6 \lambda$.
\begin{table}[h]
\caption{\label{FS_time_m_vary}
Average calculative times for circularly polarized vector vortex Bessel beams.
}
\begin{ruledtabular}
\begin{tabular}{cccc}
$m$ & $t_{2}$ (s) & $t_{1}$ (s) & $t_2 / t_1$ \\
\hline
% First line
0 & 4.2 & 0.040 & 105 \\
%End first line
%-----------------------------------------------
% Second line
1 & 4.0 & 0.038 & 105 \\
%End second line
%---------------------------------------------------
% 3rd line
2 & 4.0 & 0.037 & 108\\
%End 3rd line
%-------------------------------------------------------
% 4th line
3 & 4.0 & 0.037 & 108\\
%End 4th line
%---------------------------------------------------------
% 5th line
4 & 4.1 & 0.035 & 117 \\
%End 5th line
%---------------------------------------------------------
5 & 4.1 & 0.035 & 117 \\
10 & 4.2 & 0.033 & 127 \\
\end{tabular}
\end{ruledtabular}
\end{table}
From the table, the relation $t_2 / t_1$ is greater than $100$ for all the considered beams.
Extrapolating this to the spacial ROI considered in the previous paragraph, one obtains
the total calculative times relation
$T_2 / T_1 = (9.5 \times 10^3 / 70) \times (t_2 / t_1) \approx 1.36 \times 10^4$,
which means that in this case the single-integral algorithm will be four orders faster than
the double-integral one.

\subsubsection{Spacial-point expansion algorithm}

In case when cylindrical-coordinate scan cannot be used, azimuth-distance factorization is not
applicable directly and this advantage of the single-integral algorithm cannot be exploited.
However, as lateral coordinate scanning steps are finite (and normally small) and focal light field
components are continuous differentiable functions of distance coordinates,
we can exploit power series expansion of light field at a given distance coordinate $\rho$.
Here, we will investigate this task.
A particular (and the most widely used) form of this task can be formulated as follows:
We scan a square spacial ROI with the coordinate ranges $x = [x_1 \dots x_2]$, $y = [y_1 \dots y_2]$,
scanning steps $\Delta x = \Delta y = \Delta$, number of steps $N$ for each coordinate
and need to calculate a focal light field at each
point of the ROI $(x_i,y_j)$ according to Eq.~\eqref{FocalAmpsFin}. In such a case, distance coordinate
$\rho$ will be varied in the range $[r_1 \dots r_2]$ and take on a finite set of values:
$\rho_{ij} = \sqrt{x_i^2 + y_j^2}$.
Such a set---although being finite---will contain a number of values of the order $N^2$, which prevents
time saving based on distance-azimuth factorization in Eqs.~\eqref{FocalAmpsFin}.
So, the task should be reformulated as follows: We have a distance coordinate range $\rho = 
[r_1 \dots r_2]$ and need to calculate the amplitude-expansion vectors $\mathbf{A}_{j,\pm} (\rho,z)$
given by Eq.~\eqref{A_vector} at \textit{any} point of this range.
This is the most generalized form of the task, which can be applied not only to square ROIs
but to ROIs of any not cylindrical geometry.

To solve this task for $\rho$-dependent amplitude vectors, we should first solve it for a vector
component $A_m(\rho,z)$ given by Eq.~\eqref{AmplGener}.
So, we have a set of the distance coordinate values $S_\rho = [\rho_0, \rho_1, \rho_2 \dots \rho_M]$,
where $\rho_0 = r_1$, $\rho_M = r_2$, $\Delta \rho = (r_2 - r_1) / M$, $\rho_j = r_1 + j \Delta \rho$,
$j = [0 \dots M]$, and need to calculate $A_m$ at an arbitrary $\rho = \rho_\alpha + \delta \rho$ from this
range (the longitudinal coordinate $z$ is fixed). Here, $\rho_\alpha$ is an element of the set $S_\rho$,
whose numerical value is the closest to the target $\rho$, and the difference $\delta \rho$ can take on any
value from the range $[-\Delta \rho / 2 \dots \Delta \rho / 2]$. The target function is calculated as
\begin{equation}
\label{Am42395}
A_m(\rho,z) = A_m(\rho_\alpha,z) + \sum_{k=1}^{\infty} \frac{1}{k!} \frac{\partial^k A_m(\rho,z)}
{\partial \rho^k} \Big | _{\rho = \rho_\alpha} \delta \rho ^k.
\end{equation}
As $\rho$-dependence of this function is defined by only Bessel function $J_m ( k_1 \rho \sin \theta_1)$
[see Eq.~\eqref{AmplGener}], we have to explore the behavior of the Bessel function's derivative of an
arbitrary order and it  may be considered to depend only on $\rho$ with constant $\theta_1$ and $k_1$
to replace the partial derivative by a total one.
Exploiting, for example, Eq.~\eqref{BessFuncIntegr4237} as a Bessel function representation, one can
readily express Bessel function $k^{th}$-order derivative $J_l^{(k)} (x)$ as
\begin{equation}
\label{BessFDer237698}
J_l^{(k)} (x) = 2^{-k} \sum_{i=0}^{k} \frac{(-1)^i k!}{i! (k-i)!} J_{ l + 2 i - k }(x).
\end{equation}
Finally, Eq.~\eqref{Am42395} can be transformed into
\begin{equation}
\label{Am54689}
A_m(\rho,z) = A_m(\rho_\alpha,z) + \sum_{p=1}^{\infty} C_{m}^p(\rho_\alpha, z) \frac{(k_1 \delta \rho)^p}{p!}
\end{equation}
with
\begin{eqnarray}
\label{Cmp_compon567}
 C_{m}^p (\rho_\alpha, z) = \int \displaylimits_{0}^{\theta_{\mathrm{max}}}  K(\theta_1)  F_m (\theta_1)
T(\theta_1, z) (\sin \theta_1)^p \nonumber \\
\times J_{l(m)}^{(p)}(k_1 \rho_\alpha \sin{\theta_1}) \mathrm{d} \theta_1.
\end{eqnarray}
The multiplier $(k_1 \delta \rho)^p / p! $ decreases monotonically with increasing $p$ if $k_1 \delta \rho \le 1$
and the coefficient $C_m^p(\rho_\alpha, z)$ satisfies an inequality
\begin{equation}
|C_m^p(\rho_\alpha, z)| \le \int \displaylimits_0^{\theta_{\mathrm{max}}} |K(\theta_1)
F_m(\theta_1)T(\theta_1,z)| (\sin \theta_1)^p \mathrm{d} \theta_1.
\end{equation}
To obtain this inequality, we exploited the fact that absolute value of an integral is less than or equal to
the integral of absolute value of its integrand and the assessment
$|J_{l(m)}^{(p)}(k_1 \rho_\alpha \sin \theta_1)| \le 1$, which can be obtained from
Eq.~\eqref{BessFuncIntegr4237}.
Introducing a $p$-independent entity
\begin{equation}
\label{beta_comp}
\beta_m = \int \displaylimits_0^{\theta_{\mathrm{max}}} |K(\theta_1)
F_m(\theta_1)T(\theta_1,z)| \mathrm{d} \theta_1,
\end{equation}
we can write
\begin{equation}
|C_m^p(\rho_\alpha, z)| \le \beta_m.
\end{equation}
Now, it is possible to leave terms from $p=1$ to $p = L$ in Eq.~\eqref{Am54689},
rejecting others, and assess the rejected part, applying the negligibility criterion~\eqref{NeglCrit} as
\begin{align}
\frac{|A_m(\rho, z)|}{A_{en}} = \Bigg | \sum_{p=L+1}^{\infty} \frac{C_m^p(\rho_\alpha, z)}{A_{en}}
\frac{(k_1 \delta \rho)^p}{p!} \Bigg | \nonumber \\
\le \frac{\beta_m}{A_{en}} \frac{(k_1 \delta \rho)^{L+1}}{(L+1)!} \frac{1}{1 - (k_1 \delta \rho)/(L+2)}
\end{align}
Finally, Eq.~\eqref{Am54689} can be rewritten as
\begin{equation}
\label{SeriesPowA1236}
A_m(\rho,z) = \sum_{p=0}^{L} C_m^p (\rho_\alpha,z) \frac{(k_1 \delta \rho)^{p}}{p!}
\end{equation}
with the rejected part, the absolute value of which is assessed as
\begin{equation}
\label{NeglForDer2356}
\frac{|A_m^{rest}(\rho, z)|}{A_{en}} \le \frac{ 2 \beta_m}{A_{en}} \frac{(k_1 \delta \rho)^{L+1}}{(L+1)!}.
\end{equation}
The multiplier $2$ appeared due to the fact that $L \ge 0$ and $|k_1 \delta \rho| \le 1$.

Now, the task can be set as follows: we need to (1) divide the range of the distance coordinate
$[r_1 \dots r_2]$ into $M$ segments (which corresponds to $M+1$ points) and (2) calculate vectors
$\mathbf{A}_{j,\pm} (\rho_\alpha,z)$ and $L$ of their derivatives at each point.
So, we have $M+1$ points and $L+1$ vectors to calculate at each point and according to
Eq.~\eqref{NeglForDer2356}, the number of derivatives at a given point $\rho = \rho_\alpha$
does not depend on $\rho_\alpha$.
Although the series~\eqref{Am54689} converges at any value of $k_1 \delta \rho$,
we consider $|k_1 \delta \rho| \le 1$ to be preferrable because of the best convergence
at such values. The parameter $\delta \rho$ is expressed through the number of points $M$ as
\begin{equation}
\delta \rho = \frac{D_\rho}{2 M}
\end{equation}
with $D_\rho = r_2 - r_1$ being the length of the total distance coordinate range.
To minimize the number of vectros to be calculated, we need to solve the system
of equations
\begin{subequations}
	\label{MinimVect1254}
	\begin{gather}
	 \frac{ 2 \beta_m^{\mathrm{max}}}{A_{en}} \Bigg (\frac{k_1 D_\rho}{2} \Bigg )^{L+1} 
	\frac{M^{-(L+1)}}{(L+1)!} \le 10^{-8} \\
	 (M+1)(L+1) = \mathrm{min} \\
	 M \ge [0.5 k_1 D_\rho] + 1.
	\end{gather}
\end{subequations}
The first equation sets the negligibility criterion [see Eq.~\eqref{NeglCrit}] and the threshold is set one order
greater that that in Eq.~\eqref{NeglCrit} to prevent possible accumulative error effect; this threshold
can be corrected if necessary; $\beta_m^{\mathrm{max}}$ is the maximal-value component of the
vector $\mathbf{\beta}$ given by Eq.~\eqref{beta_comp}.
The second equation sets the minimization of the number of vectors that we need to compute
and the third one
guarantees that the expansion parameter $k_1 \delta \rho$ does not exceed unity in absolute magnitude.
To put it strictly, these equations should be considered for each vector component $m$ separately,
but we substituted a component $\beta_m$ by the maximal component  $\beta_m^{\mathrm{max}}$
because, first, it significantly simplifies the equations and, second, as entrance functions $F_m(\theta_1)$
are normally of mutually comparable magnitude, this substitution will not significantly affect the solution.
Of course, our simplification can increase the computation time and calculation precision, but not
decrease them.
In case of equal computation times for all vectors $\mathbf{C}^p(\rho,z)$ given by
Eq.~\eqref{Cmp_compon567} regardless of their order $p$, the minimization of number of vectors to
be calculated would lead to the
computation time minimization, but, in general, these times must increase with increasing $p$ because
the number of Bessel functions to be calculated is proportional to $p$ [see Eq.~\eqref{BessFDer237698}].
For this, we have to explore this problem.

We developed a Python script to compute vectors $\mathbf{C}^p(\rho,z)$, which calculates Bessel
function derivatives, directly using Eq.~\eqref{BessFDer237698}. Note that this equation allows us
to calculate derivatives of any order, in particular, $0^{th}$ order.
An investigation of the vector $\mathbf{C}^p(\rho,z)$ at $z=0$ similar to that presented in
Table~\ref{FS_time_m_vary} was carried out and the following behavior of the computation time was observed. 
$m=0$ at the entrance: $t_1=0.043 \, s$ for the $0^{th}$-order vector and increases continuously up to
$t_1 = 0.062\, s$ for the $20^{th}$-order vector; $m=3$ at the entrance: similarly, from $0.040 \, s$ up to $0.061 \, s$;
$m = 10$ at the entrance: from $0.036 \, s$ up to $0.065 \, s$.
Comparing these times for the $0^{th}$-order derivatives with those in Table~\ref{FS_time_m_vary},
we see that the use of Eq.~\eqref{BessFDer237698} by the computer program increased times by
about 3 ms, which is not crucial but indicates that the direct use of  Eq.~\eqref{BessFDer237698}
slightly delays calculations.
Next, it can be observed that for the considered beams the $20^{th}$-order derivative computation
consumes at most 2-fold greater time than that for the $0^{th}$-order one.
For the $0^{th}$- and $3^{rd}$-order entrance beams the computation times for the $20^{th}$-order
derivative is about 1.5 times greater than those for the $0^{th}$-order one, but for the $10^{th}$-order
entrance beam the similar relation amounts to about 2 times and it can be expected that
this relation will increase with the entrance
beam order increase. This is the first problem that prevents the arbitrary
derivative computation time assessment. The second possible problem
arises from the fact that entrance amplitudes can differ from the considered Bessel functions
and consume other calculative times, whic can significantly affect this assessment.
The solution to the both problems is Bessel function recursive relation
\begin{eqnarray}
\label{RecursB2}
x^2 (n-1) J_{n+2} (x) = - x^2 (n+1) J_{n-2} (x) \nonumber \\
+ 2 n (2 n^2 - x^2 -2) J_n (x),
\end{eqnarray}
which allows one to compute only two Bessel functions appearing in Eq.~\eqref{BessFDer237698}
and obtain others from them using only simple binary operations that consume
negligible times compared to a direct function computation.
The first and second  $\mathbf{A}$-vector derivatives
contain two and three Bessel function terms, respectively, and the time
assessment for them can, in principle, be extrapolated to an arbitrary derivative in light of Eq.~\eqref{RecursB2}.
We analyzed their computation times and compared
with those for $\mathbf{A}$-vector, displayed in Table~\ref{FS_time_m_vary}.
It was observed that the derivatives consume about 10-15$\%$ more time than vectors themselves.
We can suppose that in general $\theta_1$-dependent multipliers at the Bessel functions in
the integrals may be faster computable and the obtained relations may be other.
However, even in case the $\rho$-dependent Bessel function is the only mathematical expression to be calculated,
the recursive relation given by Eq.~\eqref{RecursB2} guarantees that computation time of an arbitrary
$\mathbf{A}$-vector derivative cannot exceed that of a vector itself more than 2-fold (at least,
for derivative orders less than 100).
From this, it can be assumed (for the practical use) that computation of an arbitrary $\mathbf{C}$-vector
in Eq.~\eqref{SeriesPowA1236} consumes on average time $t_c < 1.5 t_a$, where
$t_a$ is the respective $\mathbf{A}$-vector computation time and taking $t_c = 1.5 t_a$,
we can solve Eqs.~\eqref{MinimVect1254} under the  assumption of equal partial computation times to
minimize the total computation time.

Now, we can explore $xy$-scan of the ROI considered above with $x$ and $y$ varied in the range
$[-2 \lambda \dots 2 \lambda]$ with the steps $\Delta x$ and $\Delta y$ equal to $0.04 \lambda$.
For such a ROI, $D_\rho = 2 \sqrt{2} \lambda$ and $0.5 k_1 D_\rho = 2 \sqrt{2} \pi n_1 \approx
13.6$ (at $n_1 = 1.526$).
We calculated the factor $2 \beta_m^\mathrm{max} / A_{en}$ for the considered beams and
discovered that it
amounts to about $10^{-2} \dots 10^{-1}$. In general it can differ for other beams and we
explored  Eqs.~\eqref{MinimVect1254} at different thresholds $\varepsilon_{th} = 
0.5 A_{en} \varepsilon / \beta_m^\mathrm{max}$. The parameter $\varepsilon$ is set to be
equal to $10^{-8}$ in the present case, but it can be varied in general.
The following results were obtained:
at $\varepsilon_{th} = 10^{-6}$, $M=14$, $L=8$, $(L+1)(M+1) = 135$;
at $\varepsilon_{th} = 10^{-7}$, $M=14$, $L=9$, $(L+1)(M+1) = 150$;
at $\varepsilon_{th} = 10^{-8}$, $M=14$, $L=10$, $(L+1)(M+1) = 165$;
at $\varepsilon_{th} = 10^{-7}$, $M=15$, $L=10$, $(L+1)(M+1) = 176$;
at $\varepsilon_{th} = 10^{-10}$, $M=15$, $L=11$, $(L+1)(M+1) = 192$;
at $\varepsilon_{th} = 11^{-11}$, $M=15$, $L=12$, $(L+1)(M+1) = 208$.
So, the number of vectors to be calculated increases continuously from $135$ at $\varepsilon_{th} = 10^{-6}$
to $208$ at $\varepsilon_{th} = 10^{-11}$, which, first, demonstrates that it increases
pretty slowly and, second, at normal values of the parameter $2 \beta_m^\mathrm{max} / A_{en}$,
it is completely enough to calculate up to $210$ vectors to reach an acceptable precision.
As was assessed above, total times to calculate focal field inside the considered ROI for the fractional
Bessel beam~\eqref{FracBess5248} at the entrance are $T_1=1.1 \, h$ and $T_2 = 800 \, h$.
Assuming that now we need to calculate $210$ vectors instead of $70$ and average time for one
vector to be $1.5$ times greater, we obtain the corrected value of the single-integral algorithm time
$T_1^{new} = T_1 \times (210/70) \times 1.5 =4.5 T_1 \approx 5 \, h$, which is $160$ times less than
the double-integral
algorithm time $T_2 = 800 \, h$. A similar assessment for the circularly polarized vortex Bessel beams
presented in Table~\ref{FS_time_m_vary} gives $T_2 / T_1 \approx (100/1.5) \times 10^4 / 210 \approx
3.2 \times 10^3$, where $t_2 / t_1 \approx 100$ was assumed.
Thus, we conclude that the single-integral algorithm is at least two orders faster than double-integral
one even for $xy$-line-by-line scan.

\section{Conclusion}
We investigated the mathematical basis for a generalized time-effective algorithm to calculate
tight focusing of laser beams with arbitrary cross-section light vector distribution.
Focusing into an arbitrary planar microcavity is the most possible generalized
geometrical configuration which can be treated analytically and we investigated this case.
It was disclosed that a circularly polarized vortex vector beam series is an optimal
basis to decompose an entrance beam.
Based on this, a generalized single-integral algorithm was suggested.

A fractional vortex Bessel beam was treated as an example to test the algorithm.
This beam's circularly
polarized vortex vector beam representation is a logarithm-likeconditionally convergent
 infinite series, which may
diverge at some spacial points. However, focal field induced by this beam is presented
by a rapidly convergent series and to provide acceptable precision of the calculated
focal field, one needs to take from several terms to several tens of terms, depending on
the focal distance coordinate value.
As the entrance beam is represented by a slowly convergent series
with slowly decreasing coefficients and the focal series
cannot converge rapidly due to the entrance coefficients, the test results
can be readily generalized to any entrance beam.
It was shown that the suggested algorithm is, on average, about 5 times faster than the double-integral
one based on the direct use of Richards-Wolf method for a single spacial point calculation.
Similar investigation for the basis circularly polarized vortex entrance beams shows that
the single-integral algorithm is more than 100 times faster for them.

Finally, we tested the algorithm for calculations of focal fields in finite spacial regions (instead of
single-point calculations) as this task is one of the most widely applicable.
In this case, the single-integral algorithm is from three to four orders faster than the double-integral algorithm
if $xy$-scan line-by-line is not necessary. If such a scan is necessary, the single-integral algorithm
should be modified and the spacial-point expansion single-integral algorithm, its modified version, is
from two to three orders faster than the double-integral one.

\section*{Acknowledgment}
This work was supported by the Basic Research Plan of the Russian Academy of Sciences for the Period
up to 2025 (Project No. 0243-2021-0004).

%\clearpage
\bibliography{preprint_refs}

\end{document}